\documentclass[preprint,12pt]{elsarticle}

\usepackage[utf8]{inputenc}
\usepackage[T1]{fontenc}
\usepackage{graphicx}
\usepackage{amsmath,amssymb}
\usepackage{url}

\usepackage{subcaption}

\usepackage{multirow}

\usepackage{calc,xcolor}
\usepackage{color,soul}
\soulregister\cite7
\soulregister\citep7
\soulregister\citet7
\soulregister\citeauthor7
\soulregister\citeyear7
\soulregister\ref7
\soulregister\pageref7
\soulregister\fig7
\soulregister\eq7
\soulregister\eqref7
\soulregister\sect7
\soulregister\ul7
\soulregister\total7
\newcommand{\new}[1]{\hl{#1}}
\renewcommand{\new}[1]{#1}     

\makeatletter
\renewcommand*\env@matrix[1][\arraystretch]{%
  \edef\arraystretch{#1}%
  \hskip -\arraycolsep
  \let\@ifnextchar\new@ifnextchar
  \array{*\c@MaxMatrixCols c}}
\makeatother

\renewcommand{\vec}{\boldsymbol}
\newcommand{\mat}[1]{\textbf{#1}}
\newcommand{\T}{^\top}
\newcommand{\inv}{^{-1}}

\renewcommand{\eqref}[1]{Eq.~(\ref{#1})}

\newcommand{\R}{\mathbb{R}}

\newcommand{\E}{\mathbb{E}}
\newcommand{\V}{\mathbb{V}}
\newcommand{\N}{\mathcal{N}}

\newcommand{\Dexp}{\mathcal{D}}

\newcommand{\GP}{\mathcal{GP}}
\newcommand{\BigO}[1]{\mathcal{O}(#1)}

\newcommand{\dimx}{D}
\newcommand{\dimp}{D_i}
\newcommand{\Etheta}{\E_{\vec \theta|\Dexp}}

\newcommand{\Vtheta}{\V_{\vec \theta|\Dexp}}

\newcommand{\mumarg}{\breve{\mu}}

\newcommand{\Sigmamarg}{\breve{\Sigma}}
\newcommand{\mumarge}{\breve{\mu}_{(e)}}

\newcommand{\DHR}{D_{\mathrm{HR}}}
\newcommand{\DBH}{D_{\mathrm{BH}}}
\newcommand{\DBF}{D_{\mathrm{BF}}}
\newcommand{\DAW}{D_{\mathrm{AW}}}
\newcommand{\DJR}{D_{\mathrm{JR}}}

\DeclareMathOperator*{\argmax}{arg\,max}
\DeclareMathOperator*{\ei}{Ei}
\DeclareMathOperator*{\diag}{diag}

\DeclareMathOperator*{\tr}{tr}
\DeclareMathOperator*{\Div}{Div}

\newcommand{\tablespace}{\\[1.25mm]}
\newcommand\Tstrut{\rule{0pt}{2.6ex}}         
\newcommand\tstrut{\rule{0pt}{2.0ex}}         
\newcommand\Bstrut{\rule[-0.9ex]{0pt}{0pt}}   

\usepackage{listings}
\DeclareFixedFont{\ttb}{T1}{txtt}{bx}{n}{9}  
\DeclareFixedFont{\ttm}{T1}{txtt}{m}{n}{9}   
\DeclareFixedFont{\ttbs}{T1}{txtt}{bx}{n}{9} 
\DeclareFixedFont{\ttms}{T1}{txtt}{m}{n}{9}  
\usepackage{color}
\definecolor{deepblue}{rgb}{0,0,0.5}
\definecolor{deepred}{rgb}{0.6,0,0}
\definecolor{deepgreen}{rgb}{0,0.5,0}
\newcommand\pythonstyle{\lstset{
	language=Python,
	basicstyle=\ttm,
	otherkeywords={self},             
	keywordstyle=\ttb\color{black},
	emph={__init__,as},               
	emphstyle=\ttb\color{black},      
	stringstyle=\color{black},
	frame=none,                       
	showstringspaces=false,           %
	commentstyle=\ttm\color{gray}
}}
\newcommand\pythonstylesmall{\lstset{
	language=Python,
	basicstyle=\ttms,
	otherkeywords={self},             
	keywordstyle=\ttb\color{black},
	emph={__init__,as},               
	emphstyle=\ttbs\color{black},     
	stringstyle=\color{black},
	frame=none,                       
	showstringspaces=false            %
}}
\lstnewenvironment{python}[1][]{
	\pythonstyle
	\lstset{#1}
}{}
\newcommand\pythoninline[1]{{\pythonstyle\lstinline!#1!}}
\newcommand\pythonsmall[1]{{\pythonstylesmall\lstinline!#1!}}

\journal{arXiv}
\begin{document}
\begin{frontmatter}

\title{GPdoemd: A Python Package for Design of Experiments for Model Discrimination}
\author[label1]{Simon Olofsson}
\author[label2]{Lukas Hebing}
\author[label2]{Sebastian Niedenf\"uhr}
\author[label1]{\linebreak{}Marc Peter Deisenroth}
\author[label1]{Ruth Misener}
\address[label1]{Department of Computing, Imperial College London, SW7 2AZ, UK}
\address[label2]{Bayer AG, 51368 Leverkusen, Germany}
\ead{r.misener@imperial.ac.uk}

\begin{abstract}
    Model discrimination identifies a mathematical model that usefully explains and predicts a given system's behaviour. Researchers will often have several models, i.e.\ hypotheses, about an underlying system mechanism, but insufficient experimental data to discriminate between the models, i.e.\ discard inaccurate models. Given rival mathematical models and an initial experimental data set, optimal design of experiments suggests maximally informative experimental observations that maximise a design criterion weighted by prediction uncertainty. The model uncertainty requires gradients, which may not be readily available for black-box models. This paper (i) proposes a new design criterion using the Jensen-Rényi divergence, and (ii) develops a novel method replacing black-box models with Gaussian process surrogates. Using the surrogates, we marginalise out the model parameters with approximate inference. Results show these contributions working well for both classical and new test instances. We also (iii) introduce and discuss GPdoemd, the open-source implementation of the Gaussian process surrogate method.
\end{abstract}


\end{frontmatter}


\section{Introduction}
Biochemical engineering deals with noisy and uncertain processes. Modelling these processes is often difficult, and exacerbated by the difficulty of observing mechanisms and reactions on the molecular level. We can hypothesise several different mathematical models to explain a system's behaviour and run experiments to discriminate between the models. The idea is that there is a real, expensive-to-evaluate system, and $M$ rival models predicting the system behaviour. These models are effectively different hypotheses about some underlying system mechanism. We seek to discard all inaccurate model(s) with as few experiments as possible.

For analytical models, where the functional relationship can be expressed in closed form, extensive literature exists for design of experiments for model discrimination, e.g.\ \cite{HunterReiner1965}, \cite{BoxHill1967}, \cite{BuzziFerraris1983}, \cite{BuzziFerraris1984}, \cite{BuzziFerraris1990}, \cite{AspreyMacchietto2000} and \cite{Michalik2010}. The challenge is that most mathematical models for industrially relevant biological and chemical processes are neither simple nor analytical. From an optimisation point-of-view, they are often complex black boxes, e.g.\ legacy code representing large systems of partial differential equations. For these models, we can simulate the process at discrete locations, but gradient information with respect to model parameters is not readily available. The number of function evaluations needed for finite-difference gradient approximation may also be computationally prohibitive. Automatic differentiation \citep{AD_Neidinger2010, AD_Farrell2013, AD_Baydin2018} can be used to retrieve gradient information from some models, but will not work e.g.\ for models with non-smoothness and discontinuities (\citeauthor{ConnScheinbergVicenteBook}, \citeyear{ConnScheinbergVicenteBook}, pp.~3--5; \citeauthor{MartelliAmaldi2014}, \citeyear{MartelliAmaldi2014}; \citeauthor{BOUKOUVALA2016701}, \citeyear{BOUKOUVALA2016701}). These may be due to switches (\pythoninline{if}/\pythoninline{else} statements) or internal optimisation steps in the models.

Existing methods for design of experiments for black-box model discrimination utilise Monte Carlo techniques, e.g.\ \cite{Vanlier2014}, \cite{Drovandi2014}, \cite{Ryan2015} and \cite{Woods2017}. The optimal experimental design is found through exhaustive sampling of the design and model parameter spaces. This can incur a significant computational burden, especially if model evaluation is computationally slow \citep{Vanlier2014, Ryan2016}. Even for small-scale problems, the computational time to find the optimal next experiment can be on the order of days or weeks, often rivalling the time needed to carry out the actual experiment.

Design of experiments for parameter estimation has received more attention in literature than design of experiments for model discrimination \citep{Ryan2016}. Combined design criteria for parameter estimation and model discrimination exist, e.g.\ \cite{Atkinson2008} and \cite{Waterhouse2009}, but tend to perform sub-optimally for model discrimination. Therefore, we focus on design of experiments for model discrimination.

Section~\ref{sec:doemd} presents the classical method of design of experiments, with descriptions of design criteria from literature. Section~\ref{sec:DJR} proposes a novel design criterion based on the Jensen-R\'enyi divergence measure. Section~\ref{sec:gpsurrogates} describes a novel method of using Gaussian process surrogate models to perform design of experiments for discriminating black-box models. From the surrogate models, the predictive distributions can be computed and used with design criteria from literature or our proposed design criterion. Section~\ref{sec:gpdoemd} presents the open-source Python library GPdoemd, which implements the Gaussian process surrogate method. Section~\ref{sec:results} presents results from several case studies to assess the performance of our proposed design criterion based on Jensen-R\'enyi divergence and Gaussian process surrogate model method. Finally, Sections~\ref{sec:discussion} and \ref{sec:conclusions} conclude the paper. Table~\ref{tab:notation} presents the paper's notation.

\begin{table}[!t]
    \centering
    \caption{Summary of notation.}
    \label{tab:notation}
    \begin{tabular}{c l}
    \hline
    Symbol & Description \Tstrut\Bstrut\\ 
    \hline
    $\N(\cdot, \cdot)$ & Gaussian distribution
    \Tstrut\Bstrut\tablespace
    $\mat A\inv$ & Inverse of matrix $\mat A$, such that $\mat A \mat A\inv = \mat I$.
    \Tstrut\Bstrut\tablespace
    $\mat A\T$ & Transpose of matrix $\mat A$.
    \Tstrut\Bstrut\tablespace
    $|\mat A|$ & Determinant of matrix $\mat A$.
    \Tstrut\Bstrut\tablespace
    $\tr(\mat A)$ & Trace of matrix $\mat A$.
    \Tstrut\Bstrut\tablespace
    $\vec u$ & Design variable $\vec u \in \R^\dimx$.
    \Tstrut\Bstrut\tablespace
    $\dimx$ & Dimensionality of design variable space, $\vec u \in \R^\dimx$.
    \Tstrut\Bstrut\tablespace
    $\vec \theta_i$ & Parameters of model $i$ , $\vec \theta_i \in \R^{\dimp}$.
    \Tstrut\Bstrut\tablespace
    $\dimp$ & Dimensionality of model $i$'s parameter space, $\vec \theta_i \in \R^{\dimp}$.
    \Tstrut\Bstrut\tablespace
    $f_i$ & Model $i$
    \Tstrut\Bstrut\tablespace
    $M$ & Number of rival models $f_i$;\, $i = 1, \dots, M$.
    \Tstrut\Bstrut\tablespace
    $\vec f_i$ & Model $f_i$ evaluated at location $(\vec u, \vec \theta_i)$.
    \Tstrut\Bstrut\tablespace
    $f_{i,(e)}$ & Target dimension $e$ of model $f_i$; $e=1,\dots,E$.
    \Tstrut\Bstrut\tablespace
    $E$ & Number of target dimensions; $f_i:\,\R^{\dimx + \dimp} \to \R^E$.
    \Tstrut\Bstrut\tablespace
    $\vec \Sigma$ & Experimental noise covariance.
    \Tstrut\Bstrut\tablespace 
    $\Dexp$ & Set of experimental data
    \Tstrut\Bstrut\tablespace
    \hline
    \end{tabular}
\end{table}

This paper is based on and extends two conference papers. Most of Section~\ref{sec:gpsurrogates} (excluding Section~\ref{sec:sparse_gp_regression}) repeats \cite{Olofsson2018_ICML, Olofsson2018_PSE}. Tables~\ref{tab:icmlresults} and~\ref{tab:vanlierresults} and Figure~\ref{fig:vanlier}a appear in~\cite{Olofsson2018_ICML}. But Section~\ref{sec:DJR} and~\ref{sec:gpdoemd}, i.e.\ derivation of the Jensen-R\'enyi divergence design criterion and description of the GPdoemd software package, respectively, are novel contributions of this paper. Additionally, we introduce new examples to the results section.


\section{Background}
\label{sec:doemd}
The fundamental principle of experimental design for model discrimination is selecting the next experimental point where the model predictions differ most \citep{HunterReiner1965}. The task is to find an appropriate measure of this difference as a function of the model inputs. The measure of difference between model predictions should incorporate our confidence in said model predictions. Uncertainty in the model predictions comes mainly from uncertainty in the model parameters, which are tuned by fitting the model to noisy observations in the data set $\Dexp$. We assume zero-mean Gaussian distributed experimental noise with known (or upper-bounded) covariance $\vec \Sigma$. The experimental noise combines measurement noise and inherent system stochasticity. Skews in the noise distribution can often be handled, e.g.\ through a power transformation of the data \citep{BoxCoxTransform1964}.

\begin{figure}[!t]
    \centering
    \includegraphics[width=0.65\textwidth]{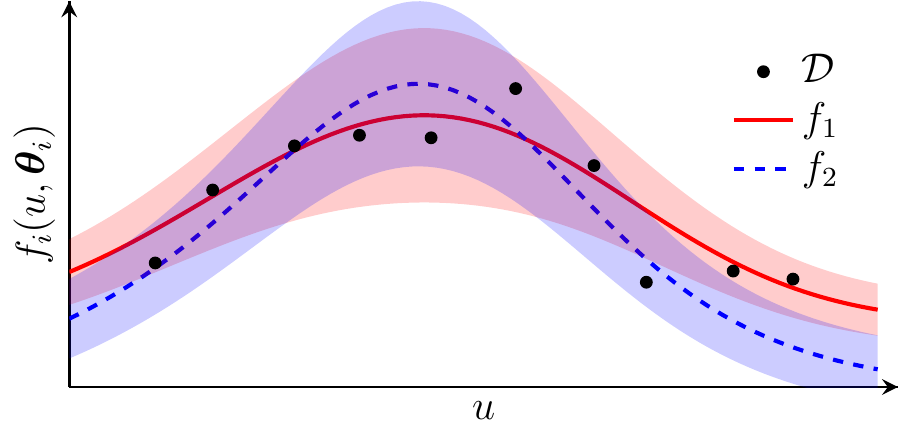}
    \caption{Observed data $\Dexp$ and two rival models $f_1$ and $f_2$ predicting the outcome of possible future experiments. The predictive means with two standard deviations are plotted.}
    \label{fig:rival_models}
\end{figure}

Let each model $i$ assume experimental observations $\vec y = y(\vec u)$ are Gaussian distributed with $\vec y \sim \N(f_i(\vec u, \vec \theta_i), \Sigma_i(\vec u))$, where $f_i(\vec u, \vec \theta_i)$ is model $i$'s predicted mean and $\Sigma_i(\vec u) = \Sigmamarg_{i}(\vec u) + \vec \Sigma$ is the sum of model and noise covariance, given design $\vec u$ and model parameter estimate $\vec \theta_i$. The experimental data set $\lbrace \vec u_n, \vec y_n \rbrace_{n=1}^N$ is denoted $\Dexp$. The model covariance $\Sigmamarg(\cdot)$ accounts for uncertainty in the model prediction due to uncertainty in the model parameter estimate $\vec \theta_i$ (see Figure~\ref{fig:rival_models}). 
To simplify notation, let $\vec f_i = f_i (\vec u, \vec \theta_i)$ and $\vec \Sigma_i = \Sigma_i(\vec u)$ denote the predictive mean and covariance, respectively.

This section first describes the classical method of computing the model covariance $\Sigmamarg_{i}(\vec u)$ through a Laplace approximation, which fits a local Gaussian to the parameter distribution $p(\vec \theta_i \,|\, \Dexp)$, centred at the maximum \emph{a posteriori} estimate $\vec \theta_i = \vec \theta_i^\ast$. Next, we describe some classical design criteria, i.e.\ measures of difference in model predictions.

\subsection{Approximating the Model Covariance $\Sigmamarg(\vec u)$}
\label{sec:paramcovar}
The predictive model covariance $\Sigmamarg(\vec u)$ accounts for variance in the model prediction $\vec f_i$ due to uncertainty in the model parameters $\vec \theta_i$. Classical literature assumes approximately linear models with respect to the model parameters near $\vec \theta_i = \vec \theta_i^\ast$ \citep{BoxHill1967, BuzziFerraris1984, PrasasSomeswaraRao1977}. Given zero-mean Gaussian distributed experimental noise with covariance $\vec \Sigma$, model $i$'s parameter estimate is approximately Gaussian distributed, $\vec \theta_i \sim \N(\vec \theta_i^\ast, \vec \Sigma_{\theta,i})$. The covariance $\vec \Sigma_{\theta,i}$ is given by the Laplace approximation:
\begin{align}
    \vec \Sigma_{\theta,i} &= 
    \left[ \sum_{n=1}^N \nabla_{\vec \theta} \vec f_i\T(\vec u_n) \vec \Sigma\inv \nabla_{\vec \theta} \vec f_i(\vec u_n) \right]\inv \,,
\end{align}
with $\nabla_\theta \vec f_i (\vec u_n) = \left. \partial f_i(\vec u_n, \vec \theta_i) / \partial \vec \theta_i \right|_{\vec \theta_i = \vec \theta_i^\ast}$ the gradient of the model prediction for design $\vec u_n \in \Dexp$ with respect to the model parameters. 
The model covariance $\Sigmamarg_{i}(\vec u)$ at a test point $\vec u$ is then given by:
\begin{align}
    \label{eq:analyticalmodelcovariance}
    \Sigmamarg_{i}(\vec u) &= 
    \nabla_{\vec \theta} \vec f_i\T(\vec u) \vec \Sigma_{\theta,i} \nabla_{\vec \theta} \vec f_i(\vec u) \,.
\end{align}
The covariance $\vec \Sigma_i$ of experimental observations according to model $i$, also taking the experimental noise covariance $\vec \Sigma$ into account, is $\vec \Sigma_i = \Sigmamarg_{i}(\vec u) + \vec \Sigma$.

\subsection{Design Criteria for Model Discrimination}
\label{sec:dcs}
The fundamental principle of sequential experimental design for model discrimination says to select the next experimental point where the model predictions differ the most \citep{HunterReiner1965}. The measure of how much the models differ is the \textit{design criterion} and the design optimisation problem maximises this design criterion.

Possibly the earliest recorded design criterion is the Mahalanobis distance between the models' predictive means, proposed by \cite{HunterReiner1965} and extended to multiple target dimensions and more than two rival models, e.g.\ by \cite{EspieMacchietto1989}:
\begin{align}
    \label{eq:DHR}
    \DHR(\vec u) = \sum_{i=1}^{M-1} \sum_{j=i+1}^M (\vec f_i - \vec f_j)\T \mat Q (\vec f_i - \vec f_j) \,,
\end{align}
where $\mat Q$ is a diagonal scaling matrix. This design criterion is popular in many practical applications, mainly because of its simplicity. \cite{AtkinsonFedorov1975} define a $T$-optimality criterion and argue that the $\DHR$ is the only design criterion that can realise $T$-optimal experimental designs.

\cite{BoxHill1967} criticise the \eqref{eq:DHR} design criterion for not considering parameter uncertainty and experimental noise, i.e.\ maximising the difference between model predictions without regard for our confidence level in the predictions' accuracy. Instead, \cite{BoxHill1967} propose measuring the information gain of an additional experimental observation $\vec y_{N}$ through the change in Shannon entropy $H_{\mathrm{S},N} = \sum_{i=1}^M \pi_{i,N} \log \pi_{i,N}$, where $\pi_{i,N} = \hat{\pi}_{i,N} / \sum_{j} \hat{\pi}_{j,N}$ are normalised model posteriors with $\hat{\pi}_{i,N} = \N(\vec y_N\,|\, \vec f_i, \vec \Sigma_i) \pi_{i,N-1}$. From this, \cite{BoxHill1967} derive a new design criterion, extended to multiple target dimensions by \cite{PrasasSomeswaraRao1977}:
\begin{align}
    \label{eq:DBH}
    \begin{split}
        \DBH(\vec u) = \sum_{i=1}^{M-1} \sum_{j=i+1}^M \pi_{i,N} \pi_{j,N} \Big\lbrace 
        &\tr ( \vec \Sigma_i \vec \Sigma_j\inv + \vec \Sigma_j \vec \Sigma_i\inv - 2 \mat I ) 
        \\&+ (\vec f_i - \vec f_j)\T ( \vec \Sigma_i\inv + \vec \Sigma_j\inv ) (\vec f_i - \vec f_j) \Big\rbrace \,.
    \end{split}
\end{align}
The \eqref{eq:DBH} design criterion is the upper bound on the expected change $\E_{\vec y_{N}}[H_{\mathrm{S},N}] - H_{\mathrm{S},N-1}$ in Shannon entropy  from the next observation $\vec y_{N}$. Experiments are conducted until $\pi_{i,N} \approx 1$ for some model $i$, or until the experimental budget is exhausted.

\cite{Meeter1970} note that it seems strange to maximise the \textit{upper} bound on the expected change in Shannon entropy rather than the \textit{lower} bound. \cite{BuzziFerraris1983} criticise the use of the normalised model posteriors $\pi_{i,N}$ in~\cite{BoxHill1967} by pointing out that observing the same values $y_{1:N}$ in different order will yield different normalised model posteriors $\pi_{i,N}$, which contradicts common statistical sense. In a series of papers \citep{BuzziFerraris1983, BuzziFerraris1984, BuzziFerraris1990}, another design criterion is proposed:
\begin{align}
    \label{eq:DBF}
    \begin{split}
        \DBF(\vec u) = \sum_{i=1}^{M-1} \sum_{j=i+1}^M &\Big\lbrace \tr \left(2 \vec \Sigma (\vec \Sigma_{i}+\vec \Sigma_{j})\inv \right)
        \\ &\quad+ (\vec f_i - \vec f_j)\T (\vec \Sigma_{i}+\vec \Sigma_{j})\inv (\vec f_i - \vec f_j) \Big\rbrace \,.
    \end{split}
\end{align}
The $\DBF$ design criterion is a heuristic based on the cross-covariance of different models' prediction errors. It can also be seen as a generalisation of $\DHR$ that incorporates parameter uncertainty~\citep[p. 5]{C_Hoffman_thesis}.

\cite{BuzziFerraris1983} and \cite{BuzziFerraris2009} note that under the null hypothesis (``the model is correct''), a model's prediction errors should be zero-mean Gaussian distributed with variance~$\vec \Sigma$. Hence, they propose using a $\chi^2$ test with $N\cdot E - D_i$ degrees of freedom to discard inaccurate models. \cite{BoxHill1967} rank models against each other and implicitly reject all but the least inaccurate model with respect to the data gathered. Using a $\chi^2$ test and discarding models that inadequately describe the observed data increases robustness against inaccurate models. Experiments are conducted until only one model remains, all models have been discarded, or the experimental budget is exhausted. For comparing two models ($M=2$), the $\DBF$ design criterion also has the interpretation that if $\DBF(\vec u)$ is not ``sufficiently larger than~1'', the models cannot be discriminated \citep{BuzziFerraris1983, BuzziFerrarisObservations}, which can be used as a stopping criterion. \cite{Schwaab2006} derive a similar design criterion to $\DBF$ consisting of the second term inside the sums in \eqref{eq:DBF} weighted by each model's prediction error $\chi^2$ probability.

\begin{figure}
    \centering
    \includegraphics[width=0.85\textwidth]{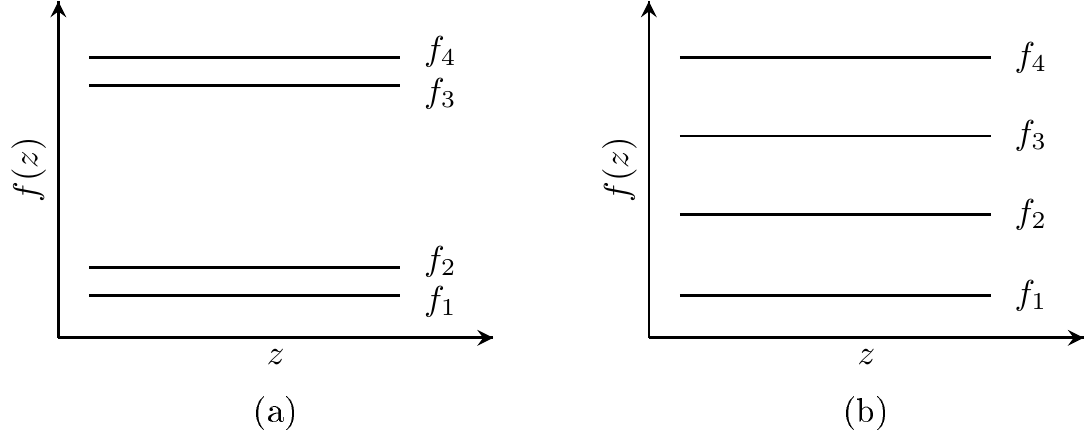}
    \caption{(a) An example of model aggregation, where $f_1$/$f_2$ and $f_3$/$f_4$ pairwise yield similar predictions. Given data, we can discriminate between groups of model pairs $f_1$-$f_2$ and $f_3$-$f_4$ but may be unable to discriminate between models intra-pair. (b) No model aggregation. Model aggregation may affect model discrimination \citep{Michalik2010}.}
    \label{fig:lumping}
\end{figure}

\cite{Michalik2010} argue that the design criteria $\DBH$ in \eqref{eq:DBH} and $\DBF$ in \eqref{eq:DBF} reward model \textit{lumping}, i.e.\ model aggregation, where the predictions of some models are similar but far apart from predictions of other model aggregations (see Figure~\ref{fig:lumping}). 
An observation at a point with model aggregation may determine whether a group of models are more accurate than another group of models, but does not discriminate between models within each of those groups. This is an interesting engineering trade-off: we may wish to identify the one best model, but significantly discriminating between model groups may be more practical than partially discriminating between many models \citep{BuzziFerrarisObservations}.

To avoid sampling at model aggregation points, \cite{Michalik2010} use Akaike's information criterion weights $w_i$ as a heuristic design criterion:
\begin{align}
    \label{eq:DAW}
    \DAW(\vec u) &= \sum_{i=1}^M w_i p(f_i) \,,
\end{align}
with $p(f_i)$ model $f_i$'s prior probability and the Akaike weights $w_i$ defined as:
\begin{align}
    w_i &= \frac{1}{\sum_{j=1}^M \exp \left( - \tfrac{1}{2} (\vec f_i - \vec f_j)\T \vec \Sigma_{i}\inv (\vec f_i - \vec f_j) + D_i - D_j \right)} \,.
\end{align}
Experiments are conducted until $w_i \approx 1$ for some model $i$ (equivalent to one model scoring a significantly higher Akaike information criterion than other models) or until the experimental budget is exhausted. The set of rival models is assumed to contain a ``good'' model, so model discrimination using $w_i$ implicitly selects the least inaccurate model. 

\section{$\DJR$: Jensen-R\'enyi Divergence Design Criterion}
\label{sec:DJR}
Model discrimination considers the difference between observed data and model predictions. Hence, \cite{HunterReiner1965} proposed designing new experiments by maximising the difference between model predictions. All three design criteria $\DBH$, $\DBF$ and $\DAW$ implicitly reward divergent model predictive distributions. This section proposes a design criterion that explicitly maximises the model predictive distributions' divergence.

The general expression for the divergence between $M$ predictive distributions $g_i(\vec u)$ for design $\vec u$ is: 
\begin{align}
    \label{eq:divergence}
    \Div[H](\vec u) = H\left( \sum_{i=1}^M \pi_i g_i(\vec u) \right) - \sum_{i=1}^M \pi_i H(g_i(\vec u)) \,,
\end{align}
where $\pi_i$ are weights associated with the corresponding models, and $H$ is some entropy measure. \cite{Vanlier2014} propose a design criterion based on the Jensen-Shannon divergence between model predictive distributions. The Jensen-Shannon divergence $\Div[H_\mathrm{S}]$ is the divergence measure corresponding to the continuous Shannon entropy $H_\mathrm{S}$, or differential entropy, defined by:
\begin{align}
    H_\mathrm{S}(G) = - \int G(\vec \gamma) \log G(\vec \gamma) \mathrm{d} \vec\gamma \,,
\end{align}
with information measured in natural units (logarithm base $e$). The expression for $\Div[H_\mathrm{S}](\vec u)$ in \eqref{eq:divergence} is intractable, even for the case of Gaussian distributions $g_i(\vec u) = \N(\vec f_i,\vec \Sigma_i)$. Hence, we can either approximate the divergence measure using Monte Carlo techniques~\citep{Vanlier2014}, or find a different entropy measure that yields a closed-form solution for the divergence of Gaussian distributions. Given the computational complexity of the former, we choose the latter option and turn to the \cite{Renyi1965} entropy measure, a generalisation of the Shannon entropy. Specifically, we look at the quadratic R\'enyi entropy $H_2$ defined as:
\begin{align}
    H_2(G) = - \log \int G(\vec \gamma)^2 \mathrm{d} \vec \gamma\,.
\end{align}
For a Gaussian distribution $g_i(\vec u)$, the quadratic R\'enyi entropy is given by:
\begin{align}
    \begin{split}
	    H_2(g_i(\vec u)) &= \tfrac{E}{2} \log (4\pi) + \tfrac{1}{2} \log \left| \vec \Sigma_{i} \right| \,.
    \end{split}
\end{align}
For a mixture of Gaussian distributions $\hat{G}=\sum_i \pi_i g_i(\vec u)$, the quadratic R\'enyi entropy is given by~\citep{Wang2009, Nielsen2012}:
\begin{align}
    \label{eq:H_2_G}
	H_{2}(\hat{G}) 
    &= - \log \sum_{i=1}^M \sum_{j=1}^M \frac{\pi_i \pi_j}{(2\pi)^\frac{E}{2}} \exp \left( - \tfrac{1}{2} \phi_{ij} (\vec u) \right) \,,
\end{align}
where, using $\tilde{\vec f}_{ij} =  \vec \Sigma_{i}\inv \vec f_{i} + \vec \Sigma_{j}\inv \vec f_{j}$, the function $\phi_{ij}(\vec u)$ is given by:
\begin{align}
    \begin{split}
        \phi_{ij} (\vec u)
        &= \vec f_{i}\T \vec \Sigma_{i}\inv \vec f_{i} + \vec f_{j}\T \vec \Sigma_{j}\inv \vec f_{j} - \tilde{\vec f}_{ij}\T \left(\vec \Sigma_{i}\inv + \vec \Sigma_{j}\inv \right)\inv \tilde{\vec f}_{ij} \\
        &\quad+ \log |\vec \Sigma_{i}| + \log |\vec \Sigma_{j}| + \log |\vec \Sigma_{i}\inv + \vec \Sigma_{j}\inv| \,.
    \end{split}
\end{align}
By noting that $\phi_{ii}(\vec u) = \log|2 \vec \Sigma_{i}|$, we rewrite the expression in \eqref{eq:H_2_G} to:
\begin{align}
    \begin{split}
        H_2(\hat{G}) &= \tfrac{E}{2} \log(2\pi) - \log \sum_{i=1}^M \left[
        \frac{\pi_i^2}{2^\frac{E}{2} | \vec \Sigma_i|^\frac{1}{2}} 
        + 2 \sum_{j=1}^{i-1} \pi_i \pi_j\exp \left( - \tfrac{1}{2} \phi_{ij} (\vec u) \right) \right] \,,
    \end{split}
\end{align}
which reduces computational effort. After these reformulations, the quadratic Jensen-R\'enyi divergence $\Div[H_2](\vec u)$ has a closed-form expression. We let:
\begin{align}
    \label{eq:DJR}
    \DJR(\vec u) = \Div[H_2](\vec u) \,,
\end{align}
denote the quadratic Jensen-R\'enyi divergence design criterion.

The weights $\pi_i$ are not uniquely defined in $\DJR$. They can be defined e.g.\ as the normalised posteriors $\pi_{i,N}$ of \citet{BoxHill1967}, the Akaike weights $w_i$ of \cite{Michalik2010}, or as $\pi_i=1$ if the model passes the $\chi^2$ test \citep{BuzziFerraris1983} and $\pi_i = 0$ otherwise. Similarly, the new $\DJR$ design criterion is not derived from any specific method of model discrimination, but from the fundamental principle of maximising model predictive divergence. Thus, the design criterion $\DJR$ is agnostic to the methods used for model weighting and discrimination. 

\subsection{Design Criteria Trade-offs}

\begin{figure}[!t]
    \centering
    \includegraphics[width=0.95\textwidth]{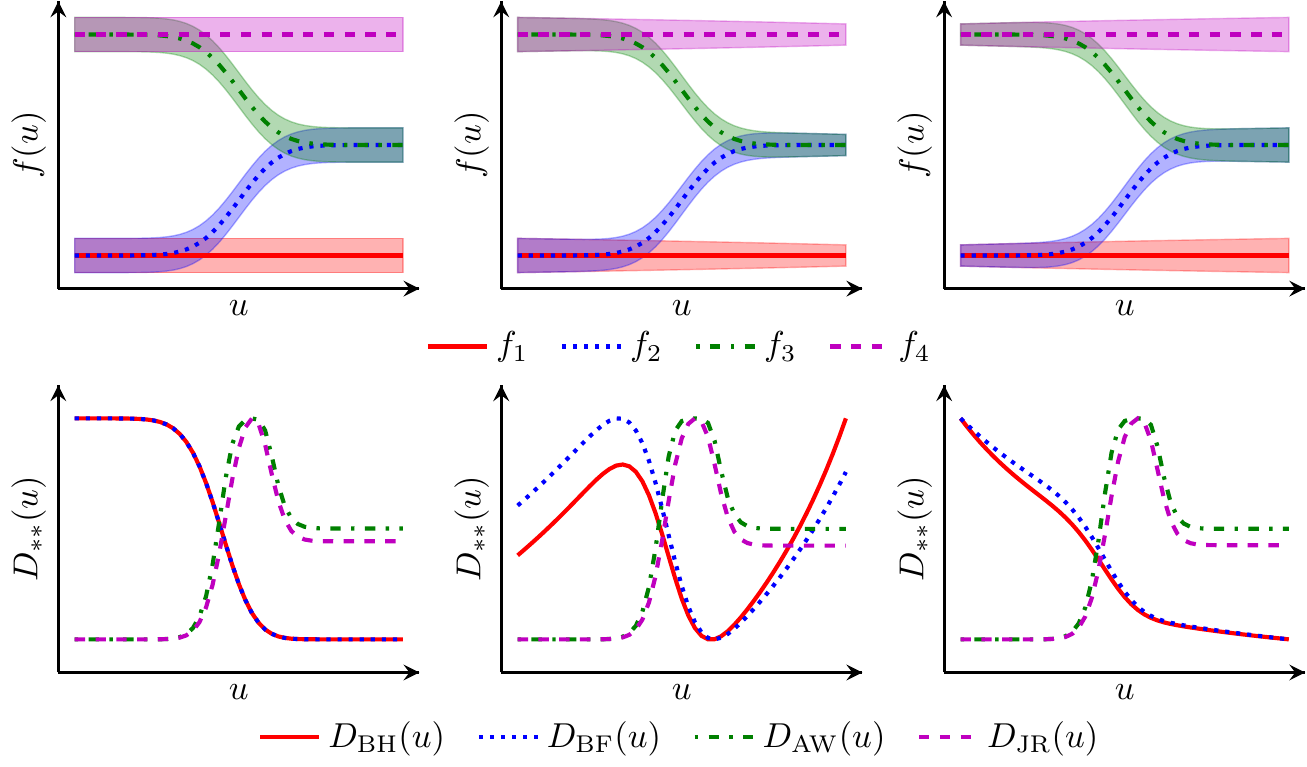}
    \caption{Comparison of discrimination criteria (bottom row) for three different sets of predictive distributions (top row): constant variance (left), linearly decreasing variance (centre) and linearly increasing variance (right).}
    \label{fig:dcs1}
\end{figure}

Figure~\ref{fig:dcs1} compares the different design criteria (bottom row) for three sets of different predictive distributions (top row) for four equally weighted models ($\forall\,i\!:\pi_i = 1/4$). The means $f_i(u)$ of the predictive distributions are the same in each plot, but the variance $\sigma_i^2(u)$ changes. The design criteria have all been normalised to lie in the range $[0,\,1]$.

In the left-most Figure~\ref{fig:dcs1} column, the variance is constant. The top plot shows a model aggregation example. The $\DBH$ and $\DBH$ criteria prefer a small $u$, where $f_1$/$f_2$ and $f_3$/$f_4$ pairwise yield identical predictions, but the divergence between e.g.\ $f_1$ and $f_3$ is large. For constant covariance, i.e.\ independent of $u$, $\DBH, \DBF \propto \DHR$. The $\DAW$ and $\DJR$ criteria, on the other hand, are maximised for medium $u$, where all model predictive distributions are divergent. In the middle Figure~\ref{fig:dcs1} column, the variance decreases linearly with $u$. In the right-most column, the variance increases linearly. In the middle and right-most columns, the change in the variance significantly impacts the $\DBH$ and $\DBF$ criteria maxima. The $\DAW$ and $\DJR$ criteria consistently favour a medium $u$, i.e.\ aim for perfect model discrimination.

\begin{figure}[!t]
    \centering
    \includegraphics[width=0.95\textwidth]{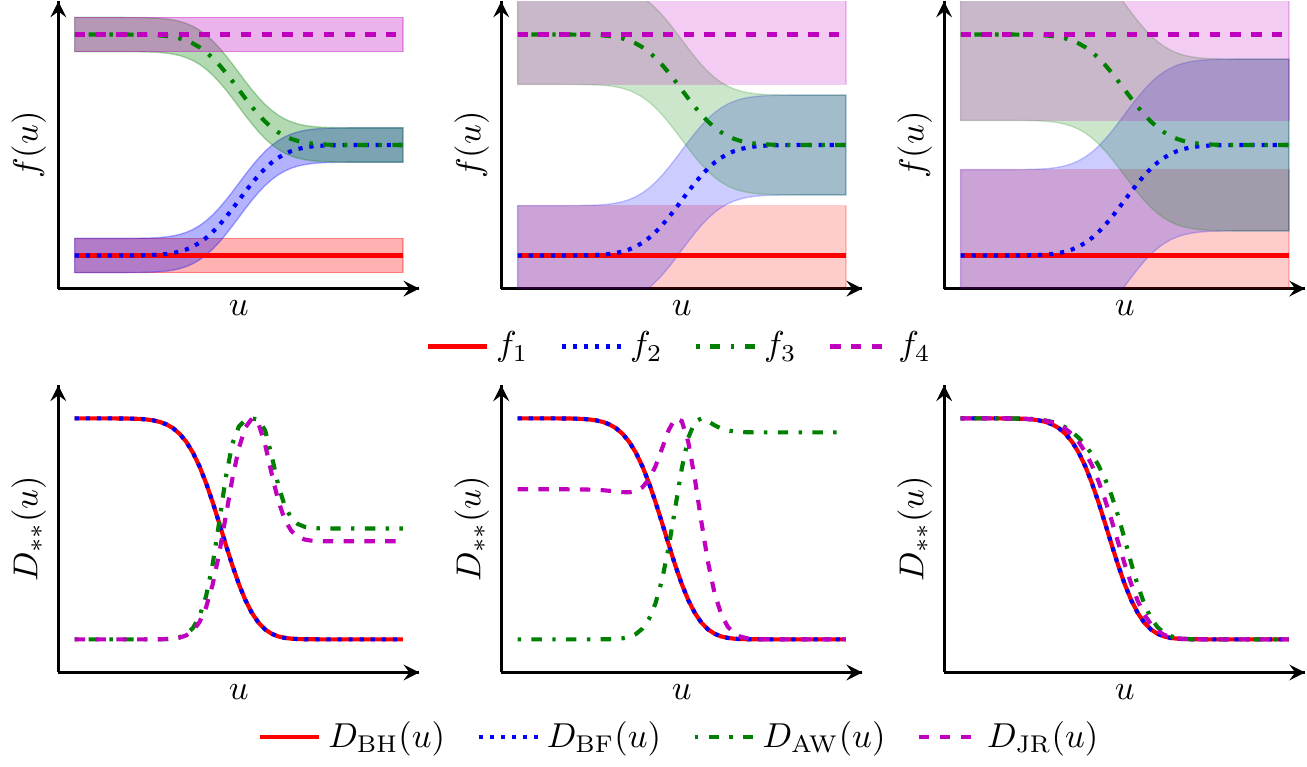}
    \caption{Comparison of discrimination criteria (bottom row) for three different sets of predictive distributions (top row) with constant covariance increasing from left to right.}
    \label{fig:dcs2}
\end{figure}

\begin{figure}[!t]
    \centering
    \includegraphics[width=0.85\textwidth]{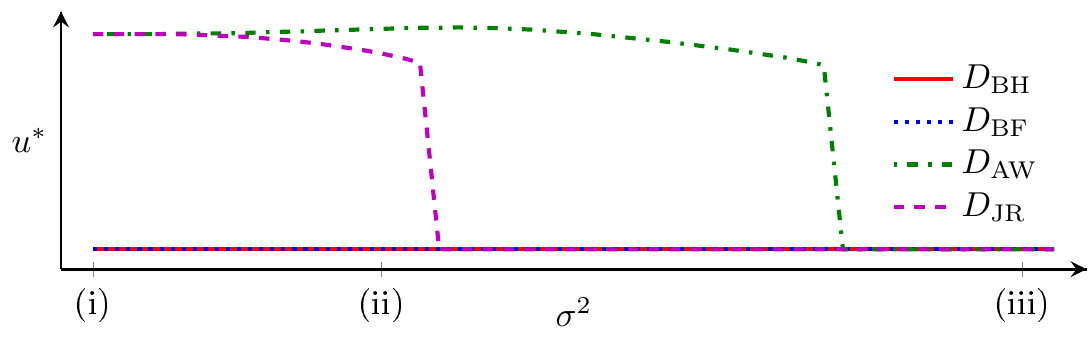}
    \caption{Optimal design $u^\ast = \argmax_u D_{\ast\ast}(u)$ when variance $\sigma^2$ increases for rival models in Figure~\ref{fig:dcs2}. Variances corresponding to the (i) left-most, (ii) centre and (iii) right-most plots of Figure~\ref{fig:dcs2} are marked on the $\sigma^2$ axis. $\DBH$ and $\DBF$ are on top of each other.}
    \label{fig:u_opt}
\end{figure}

In all three cases in Figure~\ref{fig:dcs1}, none of the predictions $f_i(u) \pm \sigma_i(u)$ overlap for medium-sized $u$. Using \citeauthor{Michalik2010}'s (\citeyear{Michalik2010}) model aggregation-based reasoning, placing the next experiment at moderate $u$ yields complete model disaggregation. However, with increasing model prediction uncertainty, the attractiveness of an experiment in the centre decreases, as shown in Figure~\ref{fig:dcs2} for the constant variance scenario.
With increasing uncertainty, the $\DJR$ criterion peak shifts to smaller $u$ before the $\DAW$ criterion peak (see Figure~\ref{fig:u_opt}). 
\cite{BuzziFerrarisObservations} argue that complete discrimination between groups of models is preferable to partial discrimination between all models.  Figure~\ref{fig:dcs1} and~\ref{fig:dcs2} show that the design criteria represent different trade-offs between the risk of partial discrimination, and the reward of complete discrimination.

The design criteria are normalised to lie in the $[0,1]$ range for each plot independently. Hence, the maximum value of the design criteria in the left-most and right-most plots in Figure~\ref{fig:dcs2} are unequal, i.e.\ all design criteria prefer small $u$ in the left-most plot over small $u$ in the right-most plot.


\section{Design of Experiments for Black-Box Model Discrimination}
\label{sec:gpsurrogates}

We now consider designing experiments for black-box model discrimination, i.e.\ discriminating models without readily available gradient information. This gradient information is needed in the classical methods of design of experiments for computing model covariance (see Section~\ref{sec:paramcovar}). 

We wish to bridge the gap between (i) classical analytical methods~\citep{BoxHill1967, BuzziFerraris1990, Michalik2010}, which are computationally cheap but with limited flexibility in the possible model types, and (ii) Monte Carlo-based methods~\citep{Vanlier2014, Ryan2015} that are flexible in model type but may be computationally prohibitive. We propose hybridising the classical and Monte Carlo-based methods: we sample the design and parameter spaces to learn surrogate models that can be incorporated into existing design and model discrimination criteria.

Surrogate models are common in applications where the original model does not easily lend itself to optimisation, e.g.~\new{\citet{PalmerRealff2002}, \citet{CaballeroGrossmann2008}}, \citet{FAHMI2012105}, \new{\citet{Boukouvala2017}, \citet{Beykal2018}}, \citet{JONES2018277}, \citet{CARPIO2018190} \new{and \citet{Yang2019}}. \new{Common surrogate models include e.g.~support vector machines~\citep{CortesVapnik1995}}. Our surrogate models are Gaussian processes (GPs), flexible regression tools common in statistical machine learning~\citep{RasmussenWilliams2006}, e.g.\ for Bayesian black-box optimisation%
~(\citeauthor{Shahriari2016}, \citeyear{Shahriari2016}; 
\citeauthor{ULMASOV20161051}, \citeyear{ULMASOV20161051}; 
\citeauthor{Mehrian2018}, \citeyear{Mehrian2018};
\citeauthor{Olofsson2019_IEEE}, \new{\citeyear{Olofsson2019_IEEE}};
\citeauthor{Babutzka2019}, \new{\citeyear{Babutzka2019}}).
GPs provide model prediction confidence bounds, and their analytical nature allows us to extend the classical analytical methods to non-analytical models. The next subsection provides background on GP regression, which infers predictive distributions for function values given function observations at training locations.  

\subsection{Gaussian Process Regression}
\begin{figure}
    \centering
    \includegraphics[width=0.85\textwidth]{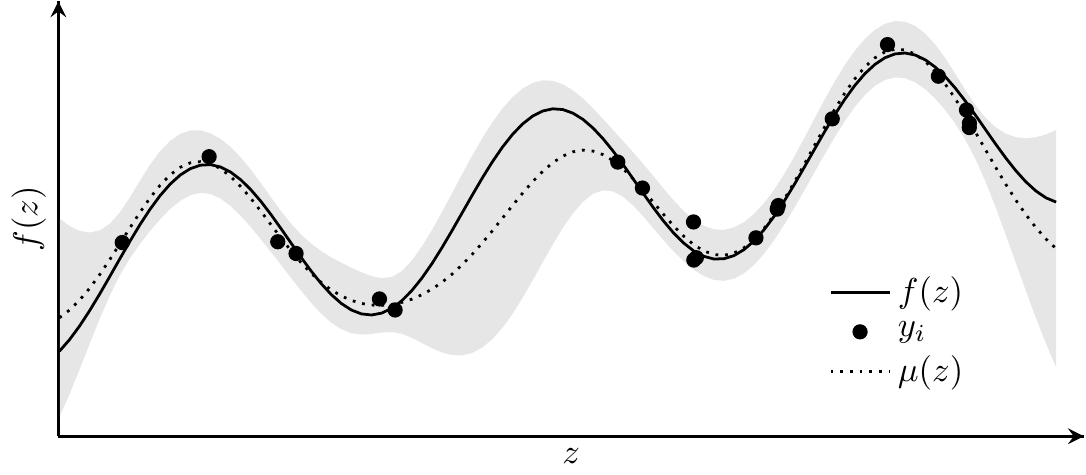}
    \caption{GP regression given noisy observations $\vec y$ (black dots) of an underlying function $f$ (solid line). The mean (dotted line) and two standard deviations (grey area) of the GP prediction are plotted.}
    \label{fig:gpr}
\end{figure}

GPs are distributions over functions. We can place a GP prior on a function $g(\vec z)$:
\begin{align}
    g \sim \GP \left(m(\cdot), k(\cdot, \cdot)\right)
\end{align}
where $m$ and $k$ are the mean function and covariance function, respectively. Formally, a GP is a \textit{collection of random variables}, any finite subset of which is \textit{jointly Gaussian distributed}~\citep{RasmussenWilliams2006}. The random variables are the values of the function $g(\cdot)$, whose joint Gaussian distribution at two locations $\vec z$ and $\vec z'$ can be written:
\begin{align}
    \begin{bmatrix} g(\vec z) \\ g(\vec z') \end{bmatrix}
    \sim \N \left(
    \begin{bmatrix} m(\vec z) \\ m(\vec z') \end{bmatrix}
    ,\,
    \begin{bmatrix} k(\vec z, \vec z) & k(\vec z, \vec z') \\ k(\vec z', \vec z) & k(\vec z', \vec z') \end{bmatrix}
    \right) \,.
\end{align}
Let $\mat Z$ and $g(\mat Z)$ denote the locations $\mat Z = [\vec z_1, \dots, \vec z_N]\T$ and corresponding function values $g(\mat Z) = [g(\vec z_1), \dots, g(\vec z_N)]\T$, and let $\mat K$ denote the Gram matrix with entries $K_{n,\ell} = k(\vec z_n, \vec z_\ell)$ and $n,\ell \in \lbrace 1,\dots,N \rbrace$. Then the function value at test point $\vec z$ and the function values at locations $\mat Z$ are jointly Gaussian distributed with:
\begin{align}
    \begin{bmatrix} g(\vec z) \\ g(\mat Z) \end{bmatrix}
    \sim \N \left(
    \begin{bmatrix} m(\vec z) \\ \vec m \end{bmatrix}
    ,\,
    \begin{bmatrix} k(\vec z, \vec z) & \vec k\T \\ \vec k & \mat K \end{bmatrix}
    \right)
\end{align}
where $\vec k = [k(\vec z, \vec z_1), \dots, k(\vec z, \vec z_N)]\T$ and $\vec m = [m(\vec z_1), \dots, m(\vec z_N)]\T$. Given observations $\vec y = g(\mat Z) + \vec \eta$, with $\vec \eta \sim \N(\vec 0, \sigma_\eta^2 \mat I)$ zero-mean Gaussian measurement noise with variance $\sigma_\eta^2$, we can compute the posterior predictive distribution at test location $\vec z$:
\begin{align}
    f(\vec z) &\sim \N(\mu(\vec z), \sigma^2(\vec z)) \,,
\end{align}
where the mean and variance are given by:
\begin{align}
    \label{eq:gp_mean_and_var}
    \begin{split}
        \mu (\vec z) &= m(\vec z) + \vec k\T (\mat K + \sigma_\eta^2\mat I )\inv (\vec y - \vec m) \,, \\
        \sigma^2 (\vec z) &= k(\vec z, \vec z) - \vec k\T (\mat K + \sigma_\eta^2\mat I )\inv \vec k \,.
    \end{split}
\end{align}
Figure~\ref{fig:gpr} illustrates GP regression given noisy observations of the underlying function. A commonly used covariance function $k$ is the radial basis function (RBF) kernel (also called the ``squared exponential'' or ``Gaussian'' kernel) with automatic relevance determination (ARD):
\begin{align}
    \label{eq:rbfardkernel}
    k(\vec z, \vec z') = \rho^2 \exp \left( -\tfrac{1}{2} (\vec z - \vec z')\T \vec \Lambda\inv (\vec z - \vec z') \right)
\end{align}
where $\rho^2$ is the signal variance and $\vec \Lambda = \diag(\lambda_{1}^2,\dots,\lambda_{D}^2)$ is a diagonal matrix of squared length scales. Covariance functions, such as the RBF-ARD kernel, are said to be stationary if they only depend on the distance $r = \| \vec z - \vec z' \|$ between $\vec z$ and $\vec z'$ and not their specific values, i.e.\ $k(\vec z, \vec z') = k(\vec z + \vec \delta, \vec z' + \vec \delta)$.

Together, $\rho^2$, $\vec \Lambda$ and $\sigma_\eta^2$ are the GP model hyperparameters. They are typically learnt by maximising the marginal likelihood $p(\vec y\,|\, \mat Z, \rho^2, \vec \Lambda, \sigma_\eta^2 )$ of the observations with respect to the hyperparameters. This can be done using gradient-based non-convex optimisation methods. The data $\mat Z$, $\vec y$ is commonly referred to as the \textit{training data}, and is used to learn the GP hyperparameters at training.

It is common to assume a zero-mean GP prior, i.e.\ $m(\cdot) \equiv 0$, for notational simplicity. If the mean is non-zero, it can be subtracted from observations and added later to predictions without affecting the result, as long as it is independent of the hyperparameters.

\subsection{Sparse Gaussian Process Regression}
\label{sec:sparse_gp_regression}
\new{The curse of dimensionality means that with increasing input dimensions, the amount of training data required to accurately model a system grows exponentially.} For large training data sets, the matrix inversion $(\mat K + \sigma_\eta^2 \mat I)\inv$ in  GP regression becomes a computational bottleneck. The computational complexity of GP training scales as $\BigO{N^3}$. Computing the predictive mean and variance in \eqref{eq:gp_mean_and_var} scales as $\BigO{N}$ and $\BigO{N^2}$, respectively (with the term $(\mat K + \sigma_\eta^2 \mat I)\inv (\vec y - \vec m)$ pre-computed).

Various methods have been proposed to reduce this computational bottleneck by sparsifying the GP regression. Sparse GP regression methods approximate the predictive distributions by selecting a smaller set of $P$ inducing points. These inducing points can either be a subset of the original training data set \citep{SmolaBartlett2001, Seeger2003} or pseduo-inputs \citep{SnelsonGhahramani2005, titsias09a}. Sparse GP regression scales as $\BigO{N P^2}$ at training, and $\BigO{P}$ and $\BigO{P^2}$ for computing the predictive mean and variance, respectively. The number $P$ is chosen as a trade-off between predictive accuracy and computational complexity.

Other methods of reducing the computational bottleneck of GP regression look at e.g.\ the spectral representation of GPs~\citep{Hensman2018} or exploiting algebraic properties of the Kronecker and Khatri-Rao tensor products on a grid of inducing inputs~\citep{EvansNair2018}. A review of sparse and variational GP regression methods---some of which can be used for training sets with billions of data points---can be found in \cite{LiuReview2018}.

\subsection{Gaussian Process Surrogate Model}
We will now show how GPs can act as surrogates for the original models for the design of experiments for model discrimination. Let's begin by studying the predictions $\vec f = f(\vec u, \vec \theta)$ of a single model $f=f_i$, with $\vec \theta = \vec \theta_i$. We wish to find the model's predictive distribution, taking uncertainty in the model parameters $\vec \theta$ into account. We place independent GP priors on each output (or target dimension) $e=1,\dots,E$ of $f$:
\begin{align}
    \label{eq:model_gp_prior}
    f_{(e)} \sim \GP(m_{(e)}(\vec x), k_{\vec u,(e)}(\vec u, \vec u') k_{\vec \theta,(e)}(\vec \theta, \vec \theta')) \,.
\end{align}
Training data is required for learning the GP hyperparameters and performing GP regression. We sample designs $\vec u_\ell$ and model parameter values $\vec \theta_\ell \sim \N(\vec \theta^\ast, \epsilon \mat I)$, for some small $\epsilon$, and produce corresponding training targets $\vec y_\ell = f(\vec u_\ell, \vec \theta_\ell)$. Let $\vec z_\ell = [\vec u_\ell\T, \vec \theta_\ell\T]\T$ denote the combined training inputs. The GP training data should not be confused with the experimental data set $\Dexp$, which is used for parameter estimation and model discrimination.

The predictive distribution $f(\vec z) \sim \N(\mu(\vec z), \Sigma_{f}(\vec z))$ at a test location $\vec z = [\vec u\T, \vec \theta\T]\T$ is given by:
\begin{align}
    \begin{split}
        \mu(\vec z) &= \left[ \mu_{(1)} (\vec z), \dots, \mu_{(E)} (\vec z) \right]\T \,,\\
        \Sigma_{f}(\vec z) &= \diag\left( \sigma_{(1)}^2 (\vec z), \dots, \sigma_{(E)}^2 (\vec z) \right) \,,
    \end{split}
\end{align}
with the expressions for $\mu_{(e)}$ and $\sigma_{(e)}^2$ given in \eqref{eq:gp_mean_and_var}. Note that the covariance function $k_{(e)}$ for each target dimension is $k_{(e)}(\vec z, \vec z') = k_{\vec u,(e)}(\vec u, \vec u') k_{\vec \theta,(e)}(\vec \theta, \vec \theta')$, as given in \eqref{eq:model_gp_prior}.

Given a model parameter distribution $p(\vec \theta\,|\,\Dexp) = \N(\vec \theta^\ast, \vec \Sigma_\theta)$, where $\vec \theta^\ast$ denotes the maximum \emph{a posteriori} parameter estimate with covariance $\vec \Sigma_\theta$, and $\Dexp$ are real, experimental data, we wish to determine the resulting marginal model predictive distribution:
\begin{align}
    \label{eq:marginal_dist}
    p \left(f (\vec u) \,\middle|\, \Dexp \right) = \int p \left( f (\vec u, \vec \theta)\,\middle|\,\vec \theta \right) p \left( \vec \theta \,\middle|\, \Dexp \right) \mathrm{d} \vec \theta \,.
\end{align}
This distribution is intractable and has to be approximated. We  approximate the marginal predictive distribution in \eqref{eq:marginal_dist} with a Gaussian distribution $\N(\mumarg(\vec u), \Sigmamarg(\vec u))$, where $\mumarg(\vec u) \approx \Etheta [ \vec \mu(\vec z) ]$ and $\Sigmamarg(\vec u) \approx \Etheta [ \Sigma_{f}(\vec z) ] + \Vtheta [ \mu(\vec z) ]$. 

We will compare first-order and second-order Taylor approximations of $\mumarg$ and $\Sigmamarg$. For the remainder of this section, we let $\vec z^{\ast} = [\vec u\T, \vec \theta^{\ast\top}]\T$ and $\Delta \vec \theta = \vec \theta - \vec \theta^\ast$, and simplify the gradient notation as $\nabla_{\vec \theta} \mu_{(e)} = \nabla_{\vec \theta} \mu_{(e)} (\vec z) |_{\vec \theta=\vec \theta^\ast}$ and $\nabla_{\vec \theta} \sigma_{(e)}^2 = \nabla_{\vec \theta} \sigma_{(e)}^2 (\vec z) |_{\vec \theta=\vec \theta^\ast}$.

\subsection{First-Order Taylor Approximation}
\label{sec:first_order_taylor}
The first-order Taylor approximations of $\mu_{(e)}(\vec z)$ and $\sigma_{(e)}^2 (\vec z)$ around a parameter value $\vec \theta^\ast$ is given by:
\begin{align}
    \begin{split}
        \mu_{(e)} (\vec z) &\approx
        \mu_{(e)} (\vec z^\ast) + \nabla_{\vec \theta} \mu_{(e)} \Delta \vec \theta \,, \\
        \sigma_{(e)}^2 (\vec z)  &\approx 
        \sigma_{(e)}^2 (\vec z^\ast) + \nabla_{\vec \theta} \sigma_{(e)}^2 \Delta \vec \theta \,.
    \end{split}
\intertext{
The expressions for $\nabla_{\vec \theta} \mu_{(e)}$ and $\nabla_{\vec \theta} \sigma_{(e)}^2$ can be found in \ref{sec:gp_gradients}. With a Gaussian model parameter distribution $\N(\vec \theta^\ast, \vec \Sigma_\theta)$, using the first-order Taylor expansions, we find:
}
    \begin{split}
        \label{eq:taylor1predDist}
        \mumarg(\vec u) &\approx \mu(\vec z^\ast) \,,\\
        \Sigmamarg(\vec u) &\approx \Sigma_{f} (\vec z^\ast) + \nabla_{\vec \theta} \vec \mu \vec \Sigma_\theta \nabla_{\vec \theta} \vec \mu\T \,.
    \end{split}
\end{align}
where $\nabla_{\vec \theta} \vec \mu \vec = [ \nabla_{\vec \theta} \mu_{(1)}\T, \dots, \nabla_{\vec \theta} \mu_{(E)}\T ]\T \in \R^{E \times \dimp}$. We see that the model covariance expression in \eqref{eq:taylor1predDist} is equivalent to the expression in \eqref{eq:analyticalmodelcovariance}, plus the added term $\Sigma_f(\vec z^\ast)$ for the uncertainty in the surrogate model. Because $\mumarg(\vec u) \approx \mu(\vec z^\ast)$ without any added terms, we may also choose to replace $\mu(\vec z^\ast)$ with the original model $f(\vec z^\ast)$. This makes the added uncertainty term $\Sigma_f(\vec z^\ast)$ in the model covariance redundant.

\subsection{Second-Order Taylor Approximation}
\label{sec:second_order_taylor}
The second-order Taylor approximations of $\mu_{(e)}(\vec z)$ and $\sigma_{(e)}^2 (\vec z)$ around a parameter value $\vec \theta^\ast$ is given by:
\begin{align}
    \begin{split}
        \mu_{(e)} (\vec z) &\approx \mu_{(e)} (\vec z^\ast) 
        + \nabla_{\vec \theta} \mu_{(e)} \Delta \vec \theta 
        + \tfrac{1}{2} \Delta \vec \theta\T \nabla_{\vec \theta}^2 \mu_{(e)} \Delta \vec \theta \,, \\
        \sigma_{(e)}^2 (\vec z) &\approx \sigma_{(e)}^2 (\vec z^\ast) 
        + \nabla_{\vec \theta} \sigma_{(e)}^2 \Delta \vec \theta
        + \frac{1}{2} \Delta \vec \theta\T \nabla_{\vec \theta}^2 \sigma_{(e)}^2 \Delta \vec \theta \,.
    \end{split}
\end{align}
The expression for $\nabla_{\vec \theta}^2 \sigma_{(e)}^2$ can be found in \ref{sec:gp_gradients}.
With a Gaussian model parameter distribution $\N(\vec \theta^\ast, \vec \Sigma_\theta)$, using the second-order Taylor expansions, we find that the marginal mean $\mumarg(\vec u)$ has elements $e=1,\dots,E$ given by:
\begin{align}
    \mumarge (\vec u) \approx \mu_{(e)} (\vec z^\ast) + \tfrac{1}{2} \tr \left( \nabla_{\vec \theta}^2 \mu_{(e)} \vec \Sigma_\theta \right) \,.
\end{align}
The corresponding covariance $\Sigmamarg(\vec u)$ is approximated with the sum of (i) the diagonal matrix $\Etheta [ \Sigma_{f} (\vec z) ] \approx \diag(q_{(1)},\dots,q_{(E)})$ with elements:
\begin{align}
    q_{(e)} &= \sigma_{(e)}^2 (\vec z^\ast) 
    + \tfrac{1}{2} \tr \left( \nabla_{\vec \theta}^2 \sigma_{(e)}^2 \vec \Sigma_\theta \right) \,,
\end{align}
and (ii) the full covariance matrix $\Vtheta [ \mu (\vec z)] \approx [q_{(e_1),(e_2)}]$ with elements:
\begin{align}
    \begin{split}
        q_{(e_1),(e_2)}
        &= \nabla_{\vec \theta} \mu_{(e_1)} \vec \Sigma_\theta \nabla_{\vec \theta} \mu_{(e_2)}\T 
        + \tfrac{1}{2} \tr \left( 
        \nabla_{\vec \theta}^2 \mu_{(e_1)} \vec \Sigma_\theta \nabla_{\vec \theta}^2 \mu_{(e_2)} \vec \Sigma_\theta 
        \right) \,,
    \end{split}
\end{align}
for $e_1,e_2 = 1, \dots, E$.

\begin{figure}
    \centering
    \includegraphics[width=\textwidth]{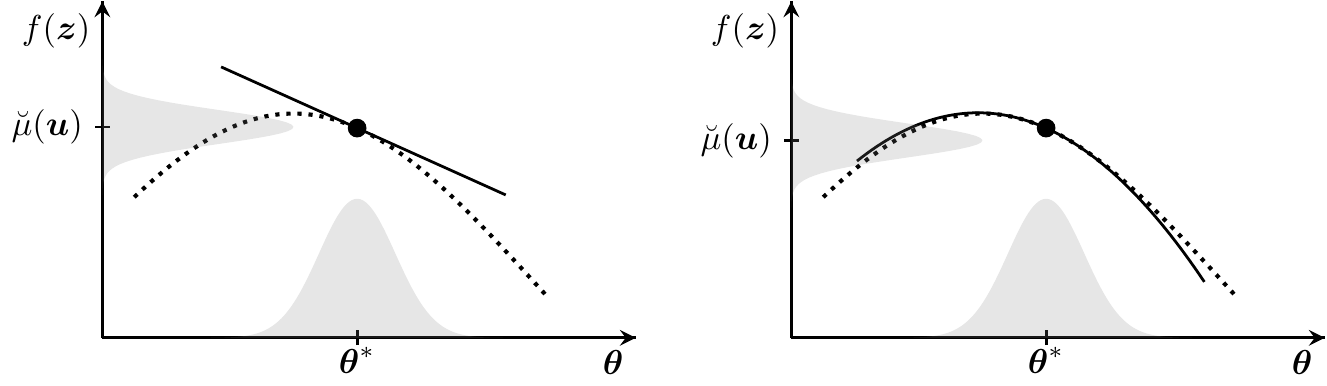}
    \caption{First- (left) and second-order (right) Taylor approximation of marginal predictive distribution (grey area on y-axis), given an input distribution (grey area on x-axis) and Taylor approximation (solid line) of the function (dotted line). The predictive distribution is shifted slightly downwards for the second-order approximation compared to the first-order one.}
    \label{fig:taylors}
\end{figure}

Figure~\ref{fig:taylors} illustrates the difference between the first- and second-order Taylor approximations. The first-order method directly correlates to the classical method of approximating the model parameter covariance and predictive distribution presented in Section~\ref{sec:paramcovar}, with $\vec f_i$ replaced with $\mumarg(\vec u)$. When using the second-order Taylor approximation of the models' predictive distributions, it is still best to use the first-order Laplace approximation of the model parameter covariance since a second-order approximation can result in a singular $\vec \Sigma_\theta\inv$.

The approximate marginalisation using first- or second-order Taylor approximations can also be used for surrogate models with inducing input sparse GP regression. The expressions for $\nabla_{\vec \theta} \mu_{(e)}$, $\nabla_{\vec \theta} \sigma_{(e)}^2$ and $\nabla_{\vec \theta}^2 \sigma_{(e)}^2$ using sparse GP regression with inducing inputs are directly equivalent to the gradient expressions for the full GP model (see \ref{sec:gp_gradients}).


\section{The GPdoemd Software Package}
\label{sec:gpdoemd}

GPdoemd\footnote{Available online at: \url{https://github.com/cog-imperial/GPdoemd}} is an open-source Python package implementing the GP surrogate method to design of experiments presented in Section~\ref{sec:gpsurrogates}. This section describes the package. Additional documentation for installing and using GPdoemd via a PDF document and Jupyter notebook demonstrations is on GitHub. GPdoemd uses functionality from the \citet{gpy2014} Python package for GP training and inference. Other dependencies are the standard numpy (v1.7-v1.15) and scipy (v0.17-v1.1) packages. GPdoemd is tested for Python version 3.4, 3.5 and 3.6. \new{There are several Python interfaces to query models written in other languages, e.g.~R or MATLAB. GPdoemd only requires point sampling of the original models in order to construct the GP surrogates.}

\subsection{Implementation}

\begin{figure}[!t]
    \centering
    \includegraphics[width=\textwidth]{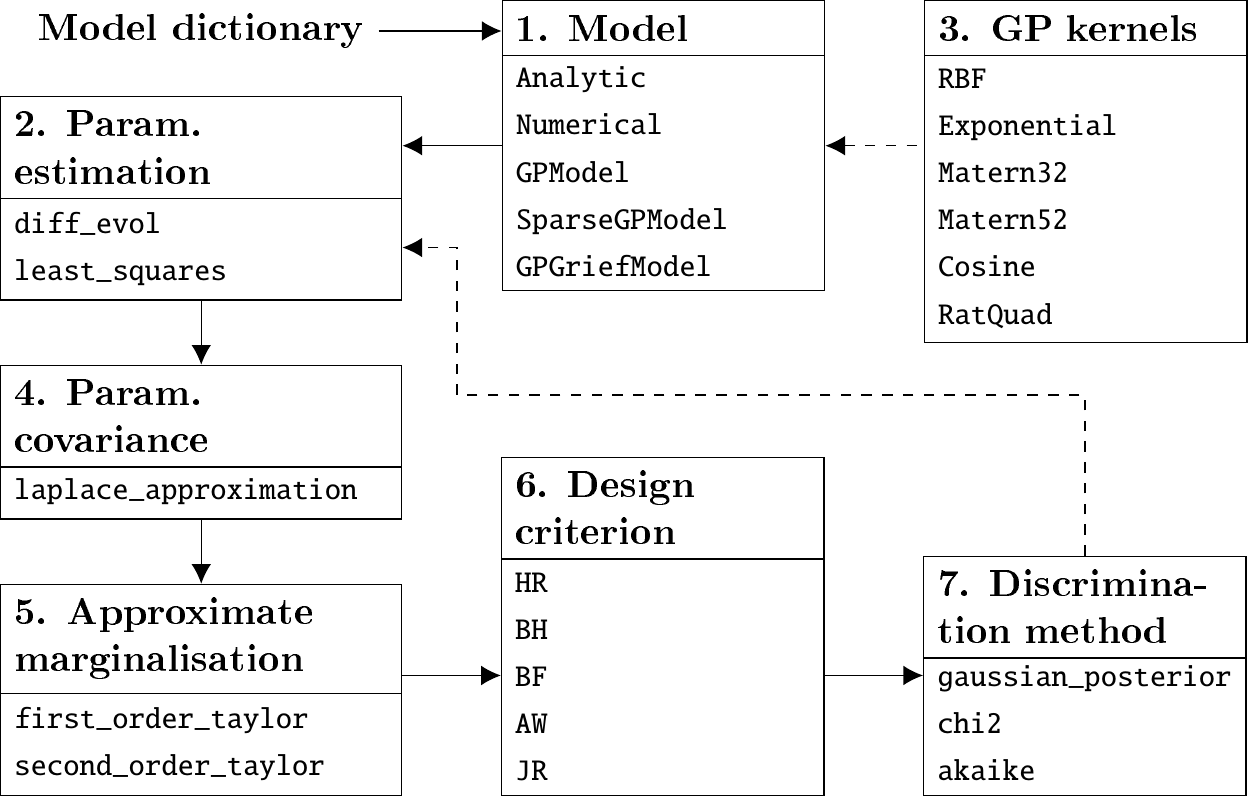}
    \caption{The modular structure of the GPdoemd open-source software package.}
    \label{fig:gpdoemd_modules}
    \vspace{-3mm}
\end{figure}

GPdoemd consists of several modules, illustrated in Figure~\ref{fig:gpdoemd_modules}, that offer a choice between different GP kernel functions, inference methods, methods to approximate the marginal predictive distributions, design criteria and model discrimination methods. The modules can easily be extended and new functions implemented and added to the GPdoemd toolbox.

GPdoemd currently comes with the Table~\ref{tab:case_studies} case studies. Researchers may try the GP surrogate method and compare the performance to competing methods for design of experiments for model discrimination. The case study \pythonsmall{mixing} was developed for GPdoemd and considers different order micro- and macrofluid models. \ref{sec:casestudymixingappendix} describes the \pythonsmall{mixing} case study.
\begin{table}
    \centering
    \footnotesize
    \begin{tabular}{l | l | c | c | c | c | l}
    	\hline 
        Name & Reference & $|\vec u|$ & $|\vec \theta_i|$ & $|\vec y|$ & $M$ & $f_i$ \Tstrut\Bstrut\\ 
        \hline
        \pythonsmall{bff1983} & \cite{BuzziFerraris1983} & 3 & 5 & 1 & 5 & A \Tstrut\Bstrut\\ 
        \pythonsmall{bffeh1984} & \cite{BuzziFerraris1984} & 2 & 4 & 2 & 4 & A \Tstrut\Bstrut\\ 
        \pythonsmall{bffc1990a} & \cite{BuzziFerraris1990} & 3 & 2--6 & 1 & 4 & A \Tstrut\Bstrut\\ 
        \pythonsmall{mixing} & -- & 3, (1) & 1 & 1 & 5 & A \Tstrut\Bstrut\\ 
        \pythonsmall{msm2010} & \cite{Michalik2010} & 3 & 1 & 1 & 10 & A \Tstrut\Bstrut\\ 
        \pythonsmall{vthr2014linear} & \cite{Vanlier2014} & 1 & 2--4 & 1 & 4 & A \Tstrut\Bstrut\\ 
        \pythonsmall{vthr2014ode} & \cite{Vanlier2014} & 3, (2) & 14 & 1 & 4 & BB \Tstrut\Bstrut\\ 
        \pythonsmall{tandogan2017} & \cite{Tandogan2017} & 4 & 8--14 & 2 & 3 & BB \Tstrut
    \end{tabular}
    \caption{GPdoemd case studies, with the number of design variables $|\vec u|$ (number of discrete variables in parenthesis), model parameters $|\vec \theta_i|$, target dimensions $|\vec y|$, rival models $M$. The last column says whether the models are analytical (A) or black boxes (BB).}
    \label{tab:case_studies}
\end{table}

\subsection{Syntax and Supported Features}
Assuming the rival models $f_i(\vec u, \vec \theta_i)$ have been proposed, GPdoemd assists in model discrimination. Figure~\ref{fig:gpdoemd_modules} illustrates the process.

\paragraph{Model type}
A model object is initialised using a Python dictionary containing the model name (\pythoninline{name}), the model function $f_i(\vec u, \vec \theta_i)$ handle (\pythoninline{call}), the design variable and model parameter dimensions $\dimx$ and $\dimp$ (\pythoninline{dim_x} and \pythoninline{dim_p}), the number of target dimensions $E$ (\pythoninline{num_outputs}), model parameter bounds (\pythoninline{p_bounds}), experimental noise (co)variance $\vec \Sigma$ (\pythoninline{meas_noise_var}), and a list of \new{the dimensions for binary design variables} (\pythoninline{binary_variables}). \new{Binary design variables are handled by creating separate GP surrogates for each binary combination.} This dictionary is passed to one of the implemented model types (Box~1 in Figure~\ref{fig:gpdoemd_modules}). \new{GPdoemd uses the GPy implementation of sparse GP regression, with variational learning of the inducing inputs~\citep{titsias09a}.} 

\paragraph{Parameter estimation}
Given experimental data \pythoninline{Ydata} for designs \pythoninline{Xdata}, GPdoemd helps find the optimal model parameter values $\vec \theta^\ast$ using prediction error minimisation (Box~2 in Figure~\ref{fig:gpdoemd_modules}): differential evolution (\pythoninline{diff_evol}) or least squares with finite difference gradient approximation (\pythoninline{least_squares}). Both \pythoninline{diff_evol} and \pythoninline{least_squares} are wrappers for \pythoninline{scipy} functions.

\paragraph{GP kernels} 
The GP surrogate models require a choice of GP kernel functions $k_x$ and $k_\theta$ for the GP prior $\GP(0, k_x k_\theta)$. GPdoemd currently supports 6 kernel functions (Box~3 in Figure~\ref{fig:gpdoemd_modules}) from the \pythoninline{GPy} package, with minor extensions.

\paragraph{Model parameter covariance}
GPdoemd assumes a Gaussian approximation $\N(\vec \theta^\ast, \vec \Sigma_\theta)$ of the model parameter distribution. GPdoemd currently implements a Laplacian approximation of $\vec \Sigma_\theta$.

\paragraph{Approximating marginal predictive distributions}
The hybrid approach approximates the marginal predictive distribution in with a Gaussian distribution. GPdoemd implements the first- and second-order Taylor approximations (Box~5 of Figure~\ref{fig:gpdoemd_modules}) of the first two moments of the models' predictive distributions in \eqref{eq:marginal_dist}.

\paragraph{Design criterion}
GPdoemd provides five different criteria (Box~6 in Figure~\ref{fig:gpdoemd_modules}) for designing the next experiment: \pythoninline{HR}~\citep{HunterReiner1965},  \pythoninline{BH}~\citep{BoxHill1967},  \pythoninline{BF}~\citep{BuzziFerraris1990} and  \pythoninline{AW}~\citep{Michalik2010} and \pythoninline{JR} (Jensen-R\'enyi divergence, this work).

\paragraph{Discrimination criterion}
GPdoemd provides three different criteria (Box~7 in Figure~\ref{fig:gpdoemd_modules}) for model discrimination: normalised Gaussian posteriors $\pi_{i,N}$~\citep{BoxHill1967}, $\chi^2$ test \citep{BuzziFerraris1983}, and the Akaike information criterion weights \citep{Michalik2010}.

\subsection{Example}
Assume we have a list \pythoninline{dlist} of model dictionaries, experimental data \pythoninline{Xdata}, \pythoninline{Ydata} with experimental noise variance \pythoninline{measvar}, and lists \pythoninline{X}, \pythoninline{P} and \pythoninline{Y} of surrogate model training data (design, model parameters and predictions, respectively). We wish to select the optimal next experiment from candidates \pythoninline{Xnew}.
\begin{python}
N = Xnew.shape[0]      # Number of test points
M = len( dlist )       # Number of rival models
E = Ydata.shape[1]     # Number of target dimensions
mu, s2 = np.zeros(( N, M, E )), np.zeros(( N, M, E, E ))
for i,d in enumerate( dlist ):
    # Initialise surrogate model
    m = GPdoemd.models.GPModel(d)
    # Estimate model parameter values
    opt_method = GPdoemd.param_estim.least_squares
    m.param_estim(Xdata, Ydata, opt_method, m.p_bounds)
    # Set-up surrogate model
    RBF = GPdoemd.kernels.RBF
    Z = np.c_[ X[i], P[i] ]
    m.gp_surrogate(Z=Z, Y=Y[i], kern_x=RBF, kern_p=RBF)
    m.gp_optimise()
    # Approximate model parameter covariance
    m.Sigma = GPdoemd.param_covar.laplace_approximation( m, Xdata )
    # Approximate marginal predictive distribution at test points
    mu[:,i], s2[:,i] = GPdoemd.marginal.taylor_first_order( m, Xnew )
# Design criterion at test points
dc = GPdoemd.design_criteria.JR(mu, s2, measvar)
# Optimal next experiment
xnext = Xnew[ np.argmax(dc) ] 
\end{python}

The newly designed experiment is executed, and \pythoninline{xnext} and the new observation added to \pythoninline{Xdata} and \pythoninline{Ydata}, respectively. If model discrimination fails, the process above is repeated in order to find the optimal next experiment.


\section{Results}
\label{sec:results}

We study the novel design criterion $\DJR$ and the GP surrogate method using four case studies:
\allowdisplaybreaks{
\begin{itemize}
    \item \textit{Ammonia synthesis}~\citep{BuzziFerraris1990}
    \item \textit{Chemical kinetic models}~\citep{BuzziFerraris1984}
    \item \textit{Mixing}~(\ref{sec:casestudymixingappendix})
    \item \textit{Biochemical networks}~\citep{Olofsson2018_ICML}
\end{itemize}
}
The first case study considers four different models for synthesis of ammonia $\mathrm{NH}_{3}$ from hydrogen $\mathrm{H}_{2}$ and nitrogen $\mathrm{N}_{2}$. There are $\dimx=3$ design variables, $\dimp \in \lbrace 2, 4, 6 \rbrace$ parameters per model, and $E=1$ observable output. Each simulation has $N_0 = 5$ initial measurements and a maximum budget of 40 new experiments. \ref{sec:casestudyoneappendix} further describes the case study. 

The second case study, further described in \ref{sec:casestudytwoappendix}, has four different chemical kinetic models. There are $\dimx=2$ design variables, $\dimp = 4$ parameters per model, and $E=2$ observable outputs. Each simulation has $N_0 = 5$ initial measurements and a maximum budget of 40 additional experiments. 

The new mixing case study is third. It studies conversion of a reactant under mixing of a fluid. There are $\dimx = 3$ design variables (one of which is binary), $\dimp = 1$ parameter per model, and $E=1$ observable output. Each simulation has $N_0 = 2$ initial measurements and a maximum budget of 20 additional experiments. \ref{sec:casestudymixingappendix} further describes the case study.

The last case study is a version of the \cite{Vanlier2014} biochemical networks case study, with models consisting of systems of ordinary differential equations \citep{Olofsson2018_ICML}. There are $\dimx=3$ design variables, $\dimp=10$ parameters per model, and $E=2$ observable outputs. Each simulation has $N_0 = 20$ initial measurements and a maximum budget of 100 additional experiments. \cite{Olofsson2018_ICML} further describe the case study.

The models in the first three case studies are analytical, i.e.\ gradient expressions are available. Section~\ref{sec:dccomparison} uses the Section~\ref{sec:paramcovar} analytical expressions for a performance comparison between the novel design criterion $\DJR$ and the classical design criteria $\DBH$, $\DBF$ and $\DAW$. Section~\ref{sec:gpperformance} compares the GP surrogate method with the analytical method it emulates. Section~\ref{sec:vanliercasestudy} shows that the GP surrogate method successfully extends the analytical method for design of experiments for model discrimination to black-box models.

Let DC and MD denote the chosen design criterion and method of model discrimination, respectively. Section~\ref{sec:dcs} describes three different methods of model discrimination:
\allowdisplaybreaks{
\begin{itemize}
    \item Normalised Gaussian posteriors $\pi_{i,N}$ with updates~\citep{BoxHill1967}. The procedure terminates when $\exists i:\, \pi_{i,N} \geq 0.999$.
    \item $\chi^2$ test~\citep{BuzziFerraris1983}, where a model $i$ is deemed inadequate and discarded if $\chi_i^2 \leq 0.01$ for $N \cdot E - \dimp$ degrees of freedom, with $N = |\Dexp|$ the number of available data points.
    \item Akaike weights $w_i$~\citep{Michalik2010}. The procedure terminates when $\exists i:\, w_{i} \geq 0.999$.
\end{itemize}
}
Alternatively, simulation terminates after reaching the maximum number of additional experiments. The Table~\ref{tab:resultmetrics} statistics are collected after each simulation. For good performance, the average A should be as low as possible. A small SE value indicates the estimated A is close to the ``true'' average. The success rate S should be close to 100\,\%. An inconclusive result (true or false negative) is preferable to a failed result (false positive), since selecting an inaccurate model can incur a large cost at a later stage (see Section~\ref{sec:discussion}).

\begin{table}[!t]
    \centering
    \begin{tabular}{l l}
        \hline
        A 
        & the average number of additional experiments required for \Tstrut \\
        & \textit{successful} model discrimination, i.e.\ identifying the correct \\
        & model as the data-generating model. \Bstrut \\
        \hline
        SE
        & the standard error of the average number A of additional \Tstrut \\
        & experiments. \Bstrut \\
        \hline
        S
        & the success rate, i.e.\ the percentage of simulations in which \Tstrut \\
        & the correct model was identified as the data-generating model. \Bstrut \\
        \hline
        F
        & the failure rate, i.e.\ the percentage of simulations in which \Tstrut \\
        & a model other than the correct model was identified as the\\
        & data-generating model. \Bstrut \\
        \hline
        I
        & the rate of inconclusive simulations, i.e.\ the percentage of \Tstrut \\
        & simulations in which more than one model--or no models in the\\
        & case of the $\chi^2$ test--remain when the experimental budget has\\
        & been exhausted. \Bstrut \\
        \hline
    \end{tabular}
    \caption{Statistics collected in the simulations.}
    \label{tab:resultmetrics}
\end{table}

\subsection{Comparison of Design Criteria Performance}
\label{sec:dccomparison}

Table~\ref{tab:table_A} compares the novel design criterion $\DJR$ to the classical design criteria $\DBH$, $\DBF$ and $\DAW$ for the ammonia synthesis case study. The comparison also includes the alternative of not optimising the design, but randomly and uniformly sampling the next experimental design, denoted $U$. We use the gradient expressions in Section~\ref{sec:paramcovar} to approximate the models' marginal predictive distributions $\N(\vec f_i, \vec \Sigma_i)$. Table~\ref{tab:table_A} shows the simulation performance statistics from 100 sets of random initial measurements.

\begin{table}[!t]
    \centering
    \vspace{1mm}
    \begin{tabular}{c | *{3}{| c c c } }
        \hline
        MD & & $\pi_{i,N}$ & & & $\chi_i^2$ & & & $w_i$ & \Tstrut\Bstrut\\
        DC      & $\DBH$ & $\DJR$ &    $U$ & $\DBF$ & $\DJR$ &    $U$ & $\DAW$ & $\DJR$ &    $U$ \Tstrut\Bstrut\\
        \hline
        A       &  20.85 &  22.24 &  34.50 &  20.56 &  21.12 &  14.50 &   7.11 &   6.61 &  21.25 \Tstrut\Bstrut\\
        SE      &   0.82 &   0.72 &   1.77 &   1.43 &   1.22 &   3.02 &   0.47 &   0.49 &   1.08 \Tstrut\Bstrut\\
        S [\%]  &     81 &     87 &      2 &     81 &     84 &     10 &    100 &    100 &     73 \Tstrut\Bstrut\\
        F [\%]  &      0 &      0 &      0 &      1 &      1 &      1 &      0 &      0 &      2 \Tstrut\Bstrut\\
        I [\%]  &     19 &     13 &     98 &     18 &     15 &     89 &      0 &      0 &     25 \Tstrut\Bstrut
    \end{tabular}
    \caption{Comparison of design criteria performance for the ammonia synthesis case study. The $\DJR$ design criterion is compared to the classical design criteria $\DBH$, $\DBF$ and $\DAW$ for their corresponding model discrimination methods (see Section~\ref{sec:dcs}). The columns $U$ uniformly sample the next experimental design rather than optimising the criterion.}
    \label{tab:table_A}
\end{table}

For this case study, the new design criterion $\DJR$ performs similarly to the classical criteria: $\DJR$ has a higher average number of additional experiments (A) than $\DBH$ and $\DBF$, but also a higher success rate (S). Compared to $\DAW$, $\DJR$ has a lower average A. In all cases, the difference between the criteria's averages A is less than the sum of their standard errors SE.

The random design selection $U$ results are a sanity check: the success rate is significantly lower while the inconclusive rate is higher. For the $\pi_{i,N}$ model discrimination method, random design selection succeeded in only 2 simulations. For the $\chi^2$ model discrimination method, the average A is lower for the random design selection than for $\DBF$ and $\DJR$, due to a low success rate: random design selection only succeeded for the easier simulations. Making an informed decision for the next experimental design is obviously beneficial to reduce the number of extra experiments needed for model discrimination.

\subsection{Performance of Gaussian Process Surrogate Method}
\label{sec:gpperformance}

We next compare the GP surrogate method of approximating the marginal predictive distribution to the analytical method (Section~\ref{sec:paramcovar}). We will show that the GP surrogate method is not significantly worse at model discrimination than the analytical method, since otherwise the GP surrogates would be ineffective for extending the analytical method to black-box models. For notational convenience, we call the GP surrogate method with first- and second-order Taylor approximations of the marginal predictive distribution GP-T1 and GP-T2, respectively. \new{In each case study, training data is generated from a grid in input space to ensure some level of space-filling.}

First, we compare the GP surrogate method to the classical analytical method on the ammonia synthesis case study~\citep{BuzziFerraris1990}. Table~\ref{tab:table_gp_ammonia} shows the performance statistics of GP-T1 and GP-T2 from simulations from 100 sets of random initial measurements. We compare these statistics to the analytical method statistics in Table~\ref{tab:table_A}.
\begin{table}[!t]
    \centering
    \begin{subtable}{0.7\linewidth}
        \begin{tabular}{c | *{3}{| c c } }
            \hline
            MD & \multicolumn{2}{|c}{$\pi_{i,N}$} & \multicolumn{2}{|c}{$\chi_i^2$} & \multicolumn{2}{|c}{$w_i$} \Tstrut\Bstrut\\
            DC      & $\DBH$ & $\DJR$ & $\DBF$ & $\DJR$ & $\DAW$ & $\DJR$ \Tstrut\Bstrut\\
            \hline
            A       &  19.72 &  21.73 &  17.28 &  17.55 &   6.73 &   6.53 \Tstrut\Bstrut\\
            SE      &   0.68 &   0.70 &   1.30 &   1.12 &   0.39 &   0.41 \Tstrut\Bstrut\\
            S [\%]  &     86 &     84 &     79 &     82 &    100 &    100 \Tstrut\Bstrut\\
            F [\%]  &      0 &      0 &      1 &      2 &      0 &      0 \Tstrut\Bstrut\\
            I [\%]  &     14 &     16 &     20 &     16 &      0 &      0 \Tstrut\Bstrut
        \end{tabular}
        \caption{GP-T1 (first-order Taylor)}
    \end{subtable}
    ~\\[3mm]
    \begin{subtable}{0.7\linewidth}
        \begin{tabular}{c | *{3}{| c c } }
            \hline
            MD & \multicolumn{2}{|c}{$\pi_{i,N}$} & \multicolumn{2}{|c}{$\chi_i^2$} & \multicolumn{2}{|c}{$w_i$} \Tstrut\Bstrut\\
            DC      & $\DBH$ & $\DJR$ & $\DBF$ & $\DJR$ & $\DAW$ & $\DJR$ \Tstrut\Bstrut\\
            \hline
            A       &   6.31 &   6.13 &  17.29 &  13.64 &   3.03 &   2.94 \Tstrut\Bstrut\\
            SE      &   0.21 &   0.24 &   1.55 &   1.73 &   0.14 &   0.12 \Tstrut\Bstrut\\
            S [\%]  &     96 &     95 &     63 &     22 &    100 &     99 \Tstrut\Bstrut\\
            F [\%]  &      0 &      0 &      0 &      2 &      0 &      1 \Tstrut\Bstrut\\
            I [\%]  &      4 &      5 &     37 &     76 &      0 &      0 \Tstrut\Bstrut
        \end{tabular}
        \caption{GP-T2 (second-order Taylor)}
    \end{subtable}%
    \caption{Performance statistics of GP surrogate method with first- (GP-T1) and second-order (GP-T2) Taylor approximation of the marginal predictive distribution for the ammonia synthesis case study. Compare to statistics for analytical method in Table~\ref{tab:table_A}.}
    \label{tab:table_gp_ammonia}
\end{table}

Table~\ref{tab:table_gp_ammonia}a shows that GP-T1 performs similarly to the analytical method. The averages A are similar, taking the sometimes relatively large standard errors SE into account. In Table~\ref{tab:table_gp_ammonia}b, GP-T2 largely produces better simulation statistics than both the analytical method and GP-T1. GP-T2 appears to produce marginal predictive distributions more beneficial for model discrimination in this case study. It may be that the second-order characteristics of GP-T2 improve the marginal predictive distribution accuracy. It may also be that the models' structure (all model parameters appear in exponents in the denominator) advantage models with the fewest model parameters ($f_1$ and $f_2$). In this case study, we generate data from model $f_1$.

\begin{table}[!t]
    \centering
    \begin{tabular}{c | *{3}{| ccc} }
        \hline
        & \multicolumn{3}{c}{Analytical} & \multicolumn{3}{|c}{GP-T1} & \multicolumn{3}{|c}{GP-T2} \Tstrut\Bstrut\\
        MD & $\pi_{i,N}$ & $\chi_i^2$ & $w_i$ & $\pi_{i,N}$ & $\chi_i^2$ & $w_i$ & $\pi_{i,N}$ & $\chi_i^2$ & $w_i$ \Tstrut\Bstrut\\
        DC & $\DBH$ & $\DBF$ & $\DAW$ & $\DBH$ & $\DBF$ & $\DAW$ & $\DBH$ & $\DBF$ & $\DAW$ \Tstrut\Bstrut\\
        \hline
        A  & 2.60 & 2.87 & 2.08 & 4.31 & 2.23 & 2.72 & 4.14 & 2.29 & 2.64 \Tstrut\Bstrut\\
        SE & 0.04 & 0.12 & 0.04 & 0.09 & 0.06 & 0.08 & 0.09 & 0.07 & 0.06 \Tstrut\Bstrut\\
        S [\%] & 86.4 & 64.2 & 62.4 & 95.6 & 47.4 & 88.6 & 96.9 & 46.6 & 90.1 \Tstrut\Bstrut\\
        F [\%] & 13.6 & 5.0 & 37.6 & 4.4 & 4.8 & 11.4 & 3.1 & 9.9 & 9.9 \Tstrut\Bstrut\\
        I [\%] & 0.0 & 30.8 & 0.0 & 0.0 & 47.8 & 0.0 & 0.0 & 43.5 & 0.0 \Tstrut\Bstrut\\
    \end{tabular}
    \caption{Performance comparison between the GP surrogate method with first- (GP-T1) and second-order (GP-T2) Taylor approximation of the marginal predictive distribution, and the analytical methods, for the chemical kinetic models case study.}
    \label{tab:icmlresults}
\end{table}

Next, we compare the GP surrogate method to the classical analytical method on the chemical kinetic models case study \citep{BuzziFerraris1984}. Table~\ref{tab:icmlresults} shows the performance statistics of GP-T1, GP-T2 and the analytical method from simulations from 500 sets of random initial measurements. In this table, GP-T1 and GP-T2 perform similarly. For the $\pi_{i,N}$ and $w_i$ model discrimination methods, the GP surrogate method has higher averages A than the analytical method, but also higher success rates and lower failure rate. For the $\chi^2$ model discrimination method, the situation reverses, though the difference in average A is smaller and for GP-T1 the failure rate is still lower. Overall, Table~\ref{tab:icmlresults} indicates that the GP surrogate method is more conservative than the classical analytical method. This conservatism may arise from the added surrogate uncertainty term $\Sigma_f$ in \eqref{eq:taylor1predDist} yielding a larger predicted variance for the surrogate than the original model.

Lastly, we compare the GP surrogate method to the classical analytical method on the mixing case study (\ref{sec:casestudymixingappendix}). The mixing case study has the option from which model to generate experimental data. Table~\ref{tab:table_gp_mixing} shows the performance statistics for GP-T1, GP-T2 and the analytical method from simulations from 100 sets of random initial measurements, with experimental data generated from model $f_3$. This case study has a high success rate and low number of additional experiments required for all discrimination criteria and methods of model discrimination.

\begin{table}[!t]
    \centering
    \begin{subtable}{0.65\linewidth}
        \begin{tabular}{c | *{3}{| c c } }
            \hline
            MD & \multicolumn{2}{|c}{$\pi_{i,N}$} & \multicolumn{2}{|c}{$\chi_i^2$} & \multicolumn{2}{|c}{$w_i$} \Tstrut\Bstrut\\
            DC      & $\DBH$ & $\DJR$ & $\DBF$ & $\DJR$ & $\DAW$ & $\DJR$ \Bstrut\\
            \hline
            A       &   4.42 &   4.25 &   2.09 &   1.30 &   2.47 &   2.38 \Tstrut\Bstrut\\
            SE      &   0.09 &   0.07 &   0.04 &   0.05 &   0.06 &   0.09 \Tstrut\Bstrut\\
            S [\%]  &    100 &    100 &    100 &    100 &    100 &    100 \Tstrut\Bstrut\\
            F [\%]  &      0 &      0 &      0 &      0 &      0 &      0 \Tstrut\Bstrut\\
            I [\%]  &      0 &      0 &      0 &      0 &      0 &      0 \Tstrut\Bstrut
        \end{tabular}
        \caption{Analytical}
    \end{subtable}
    ~\\[3mm]
    \begin{subtable}{0.65\linewidth}
        \begin{tabular}{c | *{3}{| c c } }
            \hline
            MD & \multicolumn{2}{|c}{$\pi_{i,N}$} & \multicolumn{2}{|c}{$\chi_i^2$} & \multicolumn{2}{|c}{$w_i$} \Tstrut\Bstrut\\
            DC      & $\DBH$ & $\DJR$ & $\DBF$ & $\DJR$ & $\DAW$ & $\DJR$ \Bstrut\\
            \hline
            A       &   4.65 &   4.48 &   1.78 &   1.24 &   2.61 &   2.19 \Tstrut\Bstrut\\
            SE      &   0.16 &   0.13 &   0.08 &   0.09 &   0.06 &   0.09 \Tstrut\Bstrut\\
            S [\%]  &     99 &    100 &     99 &    100 &    100 &    100 \Tstrut\Bstrut\\
            F [\%]  &      0 &      0 &      0 &      0 &      0 &      0 \Tstrut\Bstrut\\
            I [\%]  &      1 &      0 &      1 &      0 &      0 &      0 \Tstrut\Bstrut
        \end{tabular}
        \caption{GP-T1 (first-order Taylor)}
    \end{subtable}
    ~\\[3mm]
    \begin{subtable}{0.65\linewidth}
        \begin{tabular}{c | *{3}{| c c } }
            \hline
            MD & \multicolumn{2}{|c}{$\pi_{i,N}$} & \multicolumn{2}{|c}{$\chi_i^2$} & \multicolumn{2}{|c}{$w_i$} \Tstrut\Bstrut\\
            DC      & $\DBH$ & $\DJR$ & $\DBF$ & $\DJR$ & $\DAW$ & $\DJR$ \Bstrut\\
            \hline
            A       &   5.75 &   4.45 &   1.21 &   1.15 &   2.27 &   2.30 \Tstrut\Bstrut\\
            SE      &   0.16 &   0.23 &   0.06 &   0.11 &   0.05 &   0.10 \Tstrut\Bstrut\\
            S [\%]  &     99 &     95 &    100 &    100 &    100 &    100 \Tstrut\Bstrut\\
            F [\%]  &      0 &      0 &      0 &      0 &      0 &      0 \Tstrut\Bstrut\\
            I [\%]  &      1 &      5 &      0 &      0 &      0 &      0 \Tstrut\Bstrut
        \end{tabular}
        \caption{GP-T2 (second-order Taylor)}
    \end{subtable}%
    \caption{Performance comparison between the GP surrogate method with first- (GP-T1) and second-order (GP-T2) Taylor approximation of the marginal predictive distribution, and the analytical methods, for the mixing case study. Noisy observed data is generated from model $f_3$.}
    \label{tab:table_gp_mixing}
\end{table}

\begin{table}[!t]
    \centering
    \begin{subtable}{0.65\linewidth}
        \begin{tabular}{c | *{3}{| c c } }
            \hline
            MD & \multicolumn{2}{|c}{$\pi_{i,N}$} & \multicolumn{2}{|c}{$\chi_i^2$} & \multicolumn{2}{|c}{$w_i$} \Tstrut\Bstrut\\
            DC      & $\DBH$ & $\DJR$ & $\DBF$ & $\DJR$ & $\DAW$ & $\DJR$ \Bstrut\\
            \hline
            A       &  54.71 &  50.79 &  50.97 &  45.00 &  26.07 &  27.70 \Tstrut\Bstrut\\
            SE      &   2.25 &   2.08 &   4.88 &   3.96 &   1.56 &   1.68 \Tstrut\Bstrut\\
            S [\%]  &     75 &     72 &     34 &     49 &     98 &     97 \Tstrut\Bstrut\\
            F [\%]  &      0 &      0 &      0 &      1 &      2 &      2 \Tstrut\Bstrut\\
            I [\%]  &     25 &     28 &     66 &     50 &      0 &      1 \Tstrut\Bstrut\\
        \end{tabular}
        \caption{Analytical}
    \end{subtable}
    ~\\[3mm]
    \begin{subtable}{0.65\linewidth}
        \begin{tabular}{c | *{3}{| c c } }
            \hline
            MD & \multicolumn{2}{|c}{$\pi_{i,N}$} & \multicolumn{2}{|c}{$\chi_i^2$} & \multicolumn{2}{|c}{$w_i$} \Tstrut\Bstrut\\
            DC      & $\DBH$ & $\DJR$ & $\DBF$ & $\DJR$ & $\DAW$ & $\DJR$ \Bstrut\\
            \hline
            A       &  53.13 &  50.88 &  49.03 &  44.23 &  26.91 &  28.05 \Tstrut\Bstrut\\
            SE      &   1.95 &   1.96 &   4.84 &   4.23 &   1.52 &   1.79 \Tstrut\Bstrut\\
            S [\%]  &     71 &     74 &     33 &     43 &    100 &     98 \Tstrut\Bstrut\\
            F [\%]  &      0 &      0 &      2 &      1 &      0 &      2 \Tstrut\Bstrut\\
            I [\%]  &     29 &     26 &     65 &     56 &      0 &      0 \Tstrut\Bstrut\\
        \end{tabular}
        \caption{GP-T1 (first-order Taylor)}
    \end{subtable}
    ~\\[3mm]
    \begin{subtable}{0.65\linewidth}
        \begin{tabular}{c | *{3}{| c c } }
            \hline
            MD & \multicolumn{2}{|c}{$\pi_{i,N}$} & \multicolumn{2}{|c}{$\chi_i^2$} & \multicolumn{2}{|c}{$w_i$} \Tstrut\Bstrut\\
            DC      & $\DBH$ & $\DJR$ & $\DBF$ & $\DJR$ & $\DAW$ & $\DJR$ \Bstrut\\
            \hline
            A       &  75.69 &  83.05 &  60.70 &  52.86 &  12.32 &  22.96 \Tstrut\Bstrut\\
            SE      &   5.82 &   2.25 &   4.13 &  11.89 &   0.82 &   2.65 \Tstrut\Bstrut\\
            S [\%]  &     16 &     39 &     33 &      7 &     59 &     45 \Tstrut\Bstrut\\
            F [\%]  &     17 &      2 &      0 &      3 &     41 &     55 \Tstrut\Bstrut\\
            I [\%]  &     67 &     59 &     67 &     90 &      0 &      0 \Tstrut\Bstrut\\
        \end{tabular}
        \caption{GP-T2 (second-order Taylor)}
    \end{subtable}%
    \caption{Performance comparison between the GP surrogate method with first- (GP-T1) and second-order (GP-T2) Taylor approximation of the marginal predictive distribution, and the analytical methods, for the mixing case study. Noisy observed data is generated from model $f_5$.}
    \label{tab:table_gp_mixing2}
\end{table}

Table~\ref{tab:table_gp_mixing2} shows the performance statistics for GP-T1, GP-T2 and the analytical method from simulating 100 sets of random initial measurements, with experimental data generated from model $f_5$. The mixing case study with data generated from model $f_5$ is more difficult, so we increase the maximum number of additional experiments to 100. Table~\ref{tab:table_gp_mixing2} shows that GP-T1 and the analytical method perform similarly. GP-T2 performs poorly for this case study: the failure rates for the $\pi_{i,N}$ and $w_i$ methods of model discrimination are often comparable to the corresponding success rates. The averages A are higher and the success rates lower for the mixing case study with data generated from model $f_5$ instead of model $f_3$. This is due to indiscriminability between models $f_4$ and $f_5$ (further discussed in Section~\ref{sec:vanliercasestudy}).

\begin{figure}[!t]
    \centering
    \includegraphics[width=\textwidth]{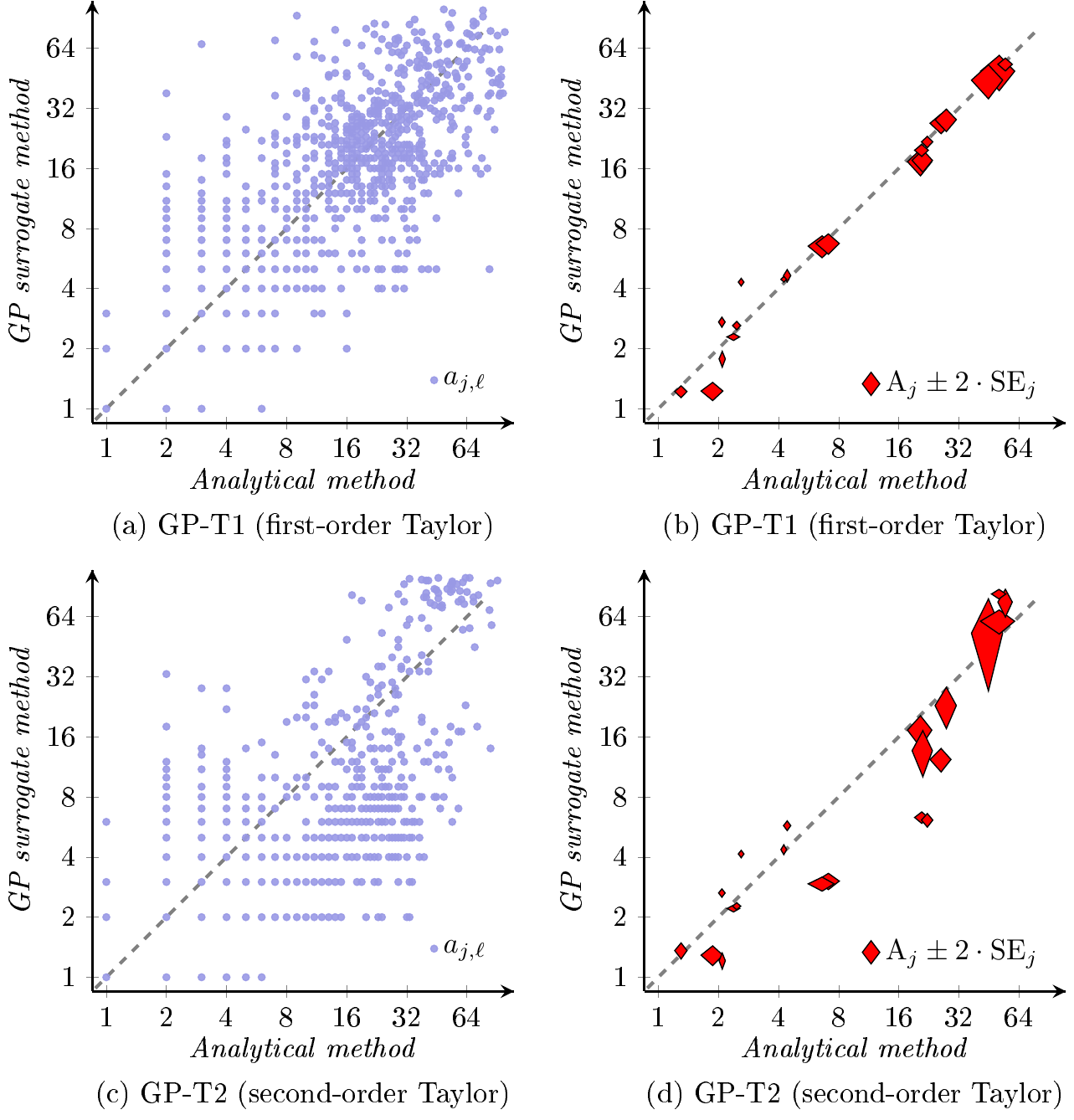}
    \caption{Comparison of the number of additional experiments $a_{j,\ell}$ needed for successful model discrimination using the analytical method or (a)-(b) GP-T1 or (c)-(d) GP-T2. Plots (a) and (c) show the outcomes of all simulation with successful model discrimination for both the analytical and GP-T1 (1828 simulations) or GP-T2 (1615 simulations). Plots (b) and (d) show the average A (with standard error SE) computed for each case study, design criterion and method of model discrimination.}
    \label{fig:Rplot}
\end{figure}

The results in Tables~\ref{tab:table_A}--\ref{tab:table_gp_mixing2} were generated for $j=1,\dots,21$ combinations of case studies, methods of model discrimination and design criteria (excluding random design). In each case study, $\ell=1,\dots,100$ random initial data sets were generated. Thus, a total of 2100 experiments were run for each of GP-T1, GP-T2 and the analytical method.
Define $a_{j,\ell}$ as the number of additional experiments needed for \textit{successful} model discrimination in combination $j$ with initial data set $\ell$ (using GP-T1, GP-T2 or the analytical method).
Simulations resulting in unsuccessful model discrimination are ignored for now. The average A and standard error SE in Table~\ref{tab:resultmetrics} are defined for combination $j$ as $\mathrm{A}_j = \mathrm{mean}\lbrace a_{j,1}, \dots, a_{j,100} \rbrace$ and $\mathrm{SE}_j = \mathrm{std}\lbrace a_{j,1}, \dots, a_{j,100} \rbrace / \sqrt{100}$. Figure~\ref{fig:Rplot} shows the outcomes of all case study simulations and compares the averages with standard errors for GP-T1, GP-T2 and the analytical method.

Figure~\ref{fig:Rplot} indicates that GP-T1 performs similarly to the analytical method. The number of additional experiments $a_{j,\ell}$ is not expected to be the same in each simulation, since we add random noise to each measurement. The similar average performance indicates that the GP surrogate method successfully emulates the analytical method, which also uses a first-order approximation.
In most case studies, GP-T2 performs as well as---or better than---GP-T1. For case studies where the GP surrogate model makes accurate predictions, a second-order approximation of the marginal predictive distributions may be more accurate than a first-order approximation. However, for the difficult mixing case study, with data generated from model $f_5$, GP-T2 performed worse. Figure~\ref{fig:Rplot} shows that the GP-T1 average performance is more similar to the analytical (first-order) method's performance than GP-T2.

\subsection{Non-Analytical Case Study}
\label{sec:vanliercasestudy}

\begin{table}[!t]
    \centering
    \begin{tabular}{c | *{3}{c} | *{3}{c} }
            \hline
            & \multicolumn{3}{c|}{\emph{4 models}} & \multicolumn{3}{c}{\emph{3 models}} \Tstrut\\
            MD& $\pi_{i,N}$ & $\chi_i^2$ & $w_i$ & $\pi$ & $\chi_i^2$ & $w_i$ \tstrut\\
            DC& $\DBH$ & $\DBF$ & $\DAW$ & $\DBH$ & $\DBF$ & $\DAW$ \tstrut\\
            \hline
            A  & 20.10 & 39.83 & 29.62 & 15.80 & 21.91 & 9.74 \Tstrut\\
            SE & 3.72 & 12.09 & 7.72 & 2.05 & 2.52 & 1.70 \tstrut\\
            S [\%] & 15.9 & 9.5 & 33.3 & 89.5 & 77.2 & 95.6 \tstrut\\
            F [\%] & 7.9 & 0.0 & 7.9 & 6.1 & 0.9 & 1.8 \tstrut\\
            I [\%] & 76.2 & 90.5 & 58.7 & 4.4 & 21.9 & 2.6 \tstrut\\
    \end{tabular}
    \caption{Results from the biochemical networks case study. With four models, we encounter model indiscriminability: Two of the models make predictions too similar to successfully discriminate between them in a majority of simulations. Experimental data is generated from one of the two models. If we remove the other model, we find that we are able to successfully perform model discrimination.}
    \label{tab:vanlierresults}
\end{table}

We wish to verify that the GP surrogate method successfully extends the classical, analytical method for design of experiment for model discrimination to situations with black-box models. The Table~\ref{tab:vanlierresults} results for 4 models show that the the success rates are significantly lower for this case study than for previous case studies. For the $\pi_{i,N}$ model discrimination method, the success rate is only twice as high as the failure rate. Rates of inconclusive results are high, despite allowing 100 additional experiments with averages A all below 50. The reason is that, in this case study, the prediction difference between models $f_1$ and $f_2$ is often smaller than the experimental noise.

\begin{figure}[!t]
    \centering
    \includegraphics[width=\textwidth]{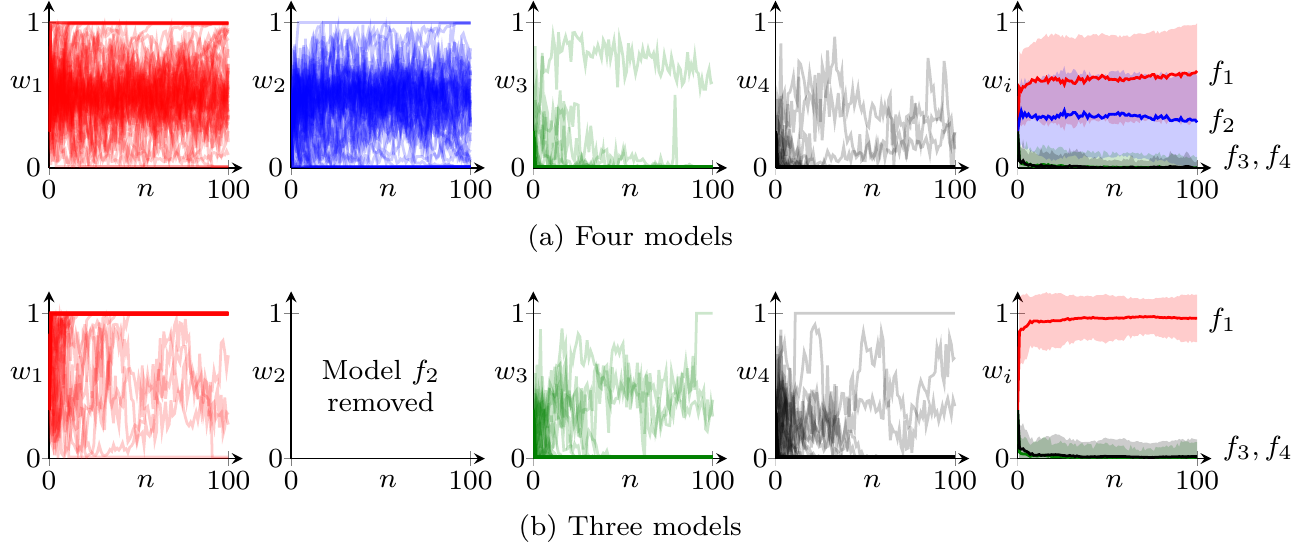}
    \caption{Results from biochemical networks case study with (a) four rival models and (b) three rival models. The first four plots from the left show the evolution of the Akaike weights $w_i$ for all simulations. The right-most plots show the averages (with one standard deviation). The plots in (a) and statistics in Table~\ref{tab:vanlierresults} indicate that model $f_1$ and $f_2$ are almost indiscriminable. When we remove model $f_2$ from the set of models, as in (b), the GP surrogate method successfully finds that $f_1$ is the data-generating model.}
    \label{fig:vanlier}
\end{figure}

For a simulation, we can examine the evolving model discrimination criterion ($\pi_{i,N}$, $\chi^2$ or $w_i$) while adding measurements. For example, Figure~\ref{fig:vanlier} shows the evolution of the Akaike weights $w_i$ for all simulations. Figure~\ref{fig:vanlier}a suggests that models $f_1$ and $f_2$ cannot be discriminated in many simulations. To verify this, we remove model $f_2$ from the set of rival models. Table~\ref{tab:vanlierresults} (3 models) and Figure~\ref{fig:vanlier}b show that removing model $f_2$ enables correct identification of model $f_1$ as the data-generating model in most simulations. 

\section{Discussion}
\label{sec:discussion}

Model discrimination is a very important problem, and optimal design of experiments forms part of the solution. For example, model discrimination is crucial in developing new drugs. \cite{DiMasi2016} estimate the average pre-approval R\&D cost for new drugs to \$2.56B, of which \$1.1B is spent in the pre-clinical stage (in 2013 US dollars). Successful model discrimination early in the drug development may lower costs, whereas inaccurate models passing the pre-clinical stage can incur significant costs \citep{ScannelBosley2016, Plenge2016}. Hence, inaccurate models should be discarded as early as possible in the drug development process. Model discrimination is a major hurdle e.g.\ in \emph{in silico} pharmacokinetics \citep{heller2018technologies}.

The optimal design of experiments for model discrimination literature has focused on either classical analytical approaches or Monte Carlo-based approaches. The former is computationally cheap but limited in the model structures and approximations it can accommodate, whereas the latter is flexible and accurate but may be computationally expensive. Hence, there is a trade-off between flexibility, accuracy and computational speed. Our GP surrogate method adds another alternative: we can accommodate non-analytical models as easily as analytical models in a computationally inexpensive way. A commonly-observed limitation of GP surrogates is the scaling to the large training data sets required to accurately emulate models with high-dimensional design and parameter spaces. As mitigation, our surrogate approach with approximate marginalisation of model parameters accommodates sparse GP surrogate models, which increases the maximum number of input dimensions the surrogate models can accommodate. \new{Because different users might face different restrictions on computational time (for generating training data and training the GP surrogates), and because different models' behaviour can differ significantly, we do not attempt to provide general guidelines for how best to generate training data. The maximum number of input dimensions the GP surrogates can accommodate depends on how much training data is required to accurately replicate the original models, which in turn depends on how sensitive each target dimension is to the model inputs.}

Beyond the GP surrogate method, we have derived the novel design criterion $\DJR$ and discussed trade-offs between different experimental designs. Investigating these trade-offs is enabled by our GPdoemd toolbox. In our experiments, the different design criteria sometimes perform similarly and (in those cases) the model discrimination method has a large impact on the success rate or the number of additional experiments required. GPdoemd allows switching between these discrimination methods. GPdoemd could be further extended to alternative model discrimination criteria including the Minimum Message Length \citep{WallaceBoulton1968}, the Bayesian Information Criterion \citep{Schwarz1978} or the Widely Applicable (Bayesian) Information Criterion \citep{Watanabe2010,Watanabe2013}. Another important consideration is parameter estimation: We may discard the data-generating model due to poor parameter estimation. GPdoemd could be extended to incorporate global parameter estimation.

Model indiscriminability is a major hurdle for model discrimination. Parametric models may be very flexible, spanning a large part of the target space depending on the specific model parameter values. \new{Useful stopping criteria for design of experiments, e.g.~\citeauthor{BuzziFerraris1983}'s (\citeyear{BuzziFerraris1983}) criterion $\DBF(\vec u) > 1$ mentioned in Section~\ref{sec:dcs}, are difficult to come by.} In practice, we may need to rethink the experimental set-up to reduce measurement noise or add new system inputs or target dimensions. Another option is to analyse the physical meaning of the model parameters with the goal of tightening the bounds on the allowed model parameter values. The smaller design space should in turn reduce the target space spanned by the model predictions.

\new{This paper has addressed sequential design of experiments, but engineers researchers may want to design several new experiments to run in parallel. This is called \emph{batch optimisation} in the Bayesian optimisation literature~\citep{Gonzalez2016}. \citet{Galvanin2006, Galvanin2007} and \citet{Bazil2012} have studied design of parallel experiments for parameter estimation, but there have been fewer contributions on design of parallel experiments for model discrimination. A simple heuristic is to design new experiments in a sequential fashion while penalising new experiments in the vicinity of experiments already added to the next batch.}

The Section~\ref{sec:results} experiments are not in every way representative of model discrimination in a real setting. The large-scale tests of the $\DJR$ design criterion and GP surrogate method require fixed thresholds for when model discrimination (using $\pi_{i,N}$, $\chi^2$ or $w_i$) rejects or selects a model. Our experiments set the ``winning'' threshold for $\pi_{i,N}$ and $w_i$ to 99.9\%, which is arguably high. But a high threshold also decreases the failure risk. In practice, an engineer may select a model maintaining a probability score, e.g.\ of $w_i=0.98$ over multiple experiments. Regardless of the precise threshold value, the results in Section~\ref{sec:results} may indicate the methods' relative performance in a real setting. 

This paper considers models of the type $\vec y = f(\vec u, \vec \theta) + \vec \epsilon$. Many processes for which we need to design experiments for model discrimination are time-dependent, and best described by state-space models. \cite{CheongManchester2014a,CheongManchester2014b} look at model discrimination and fault detection in dynamic systems, but do not take model parameter uncertainty into account. Neither do \cite{SkandaLebiedz2010,SkandaLebiedz2013}. Ignoring parameter uncertainty may lead to overconfidence in the models' accuracy. \cite{ChenAsprey2003} study non-linear state-space models with model parameter uncertainty. They use a classical approach of estimating the model parameter uncertainty using analytical gradient information. \cite{Georgakis2013} uses a data-driven approach to design dynamic experiments for model parameter estimation. We believe that our GP surrogate method can be extended to design of experiments for discrimination of rival state-space models. \cite{KoFox2009} and \cite{deisenroth2009} use GPs to learn non-linear dynamics, and propagate uncertainty in the system state to the model prediction.


\section{Conclusions}
\label{sec:conclusions}
Design of experiments for black-box model discrimination is a difficult but important problem. Our novel method, hybridising the classical and Monte Carlo-based approaches using GP surrogate models, performs similarly to the analytical approach on classical case studies and several orders of magnitude faster than existing black-box approaches. It allows flexibility with regards to the structure and software implementations of the underlying models. Limitations lie in scaling the GPs to the large training data sets required to accurately emulate models with high-dimensional design and parameter spaces. 

The GP surrogate method has been implemented in the open-source GPdoemd Python package and made available on GitHub. The GPdoemd package allows researchers and engineers to implement design of experiments for model discrimination and includes methods for approximating marginal predictive distributions, design of experiments, and model discrimination. For methodology experts wishing to develop new model discrimination approaches, GPdoemd also includes a standard set of case studies.


\section*{Acknowledgements}
\noindent
This work has received funding from the European Union's Horizon 2020 research and innovation programme under the Marie Sk\l{}odowska-Curie grant agreement no.\ 675251, and an EPSRC Research Fellowship to R.M.\ (EP/P016871/1).


\appendix

\section{Gradients of Gaussian Process Prediction}
\label{sec:gp_gradients}
The predictive mean $\mu(\vec z)$ and variance $\sigma^2(\vec z)$ are easily differentiable with respect to the input $\vec z$. The three gradients used in this paper are:
\begin{align}
    \frac{\partial \mu(\vec z)}{\partial \vec z} 
    &= (\vec y - \vec m)\T (\mat K + \sigma_\eta^2\mat I )\inv \frac{\partial \vec k}{\partial \vec z} \,, \\
    \frac{\partial^2 \mu(\vec z)}{\partial z_i \partial z_j} 
    &= (\vec y - \vec m)\T (\mat K + \sigma_\eta^2\mat I )\inv \frac{\partial^2 \vec k}{\partial z_i \partial z_j} \,, \\
    \frac{\partial^2 \sigma^2(\vec z)}{\partial z_i \partial z_j} 
    &= -2 \left[ \vec k (\mat K + \sigma_\eta^2\mat I )\inv \frac{\partial^2 \vec k}{\partial z_i \partial z_j} + \frac{\partial \vec k}{\partial z_i}\T (\mat K + \sigma_\eta^2 \mat I)\inv \frac{\partial \vec k}{\partial z_j} \right] \,. 
\end{align}
The most common covariance functions $k$ are also differentiable, e.g.\ the RBF-ARD kernel $k(\vec z', \vec z) = \rho^2 \exp(-\tfrac{1}{2}(\vec z' - \vec z)\T \vec \Lambda\inv(\vec z' - \vec z))$:
\begin{align}
    \frac{\partial k(\vec z', \vec z)}{\partial \vec z} 
    &= k(\vec z', \vec z) (\vec z' - \vec z)\T \vec \Lambda\inv \,, \\
    \frac{\partial^2 k(\vec z', \vec z)}{\partial \vec z \partial \vec z\T} 
    &= k(\vec z', \vec z) \vec \Lambda\inv \left( (\vec z' - \vec z) (\vec z' - \vec z)\T - \mat I \right)
\end{align}
Likewise, expressions can be derived for the gradients of the mean and variance with respect to the hyperparameters, gradients that are used for gradient-based learning of the hyperparameters during training.
For sparse GP regression using inducing inputs, these gradient expressions still hold, swapping the term $(\mat K + \sigma_\eta^2\mat I )\inv$ with the appropriate sparse GP equivalent.

\section{Case Study: Mixing}
\label{sec:casestudymixingappendix}
This case study considers a single fluid containing a reactant mixing during reaction. We wish to learn (i) whether it is a zero-, first- or second-order reaction, and (ii) whether mixing occurs on the microscopic or macroscopic level~\citep[ch.~16]{Levenspiel1999}. We can run experiments in a plug flow reactor (PFR) or continuous stirred-tank reactor (CSTR) and observe the reactant conversion rate. We assume ideal reactors. The rival models $f_i$ are given in Table~\ref{tab:MicroMacroModels}, where $\ei(x) = \int_{-x}^\infty t^{-1}\exp(-t) \mathrm{d}t$ is the exponential integral.\\[-6mm]
\begin{table}[!ht]
    \footnotesize
    \centering
    \begin{tabular}{ c | c | c | l | l}
        \textbf{Order} & \multicolumn{1}{c|}{$\boldsymbol R$} & \textbf{Mixing} & \multicolumn{1}{c|}{\textbf{PFR}} & \multicolumn{1}{c}{\textbf{CSTR}} \Tstrut\Bstrut\\
        \hline
        \multirow{2}{*}{$0^\mathrm{th}$} & 
        \multirow{2}{*}{$\theta u_1 / u_2$} &
        Micro &
        $f_1 = f_2 = 1 - R$ &
        $f_1 = 1 - R$ ($f_1 = 0$ if $R \geq 1$) \Tstrut\Bstrut\\ \cline{3-3}\cline{5-5}
        & & Macro & ($f_1=f_2=0$ if $R \geq 1$) &
        $f_2 = 1 - R + R\exp(-1/R)$ \Tstrut\Bstrut\\ \hline
        \multirow{2}{*}{$1^\mathrm{st}$} & 
        \multirow{2}{*}{$\theta u_1$} & 
        Micro &
        \multirow{2}{*}{$f_3 = \exp(-R)$} &
        \multirow{2}{*}{$f_3 = 1/(1+R)$} \Tstrut\Bstrut\\ \cline{3-3}
        & & Macro & &
        \Tstrut\Bstrut\\ \hline
        \multirow{2}{*}{$2^\mathrm{nd}$} & 
        \multirow{2}{*}{$\theta u_1 u_2$} & 
        Micro &
        \multirow{2}{*}{$f_4 = f_5 = 1/(1+R)$} &
        $f_4 = \tfrac{1}{2R} \left( -1 + \sqrt{1+4R} \right)$ \Tstrut\Bstrut\\[0.95mm] \cline{3-3}\cline{5-5}
        & & Macro & &
        $f_5 = \tfrac{1}{R}\exp(1/R)\ei(1/R) $ \Tstrut\Bstrut\\
    \end{tabular}
    \caption{Conversion rate models for a reaction in a micro- or macrofluid during mixing in an ideal PFR or CSTR reactor~\citep[p.~356]{Levenspiel1999}}
    \label{tab:MicroMacroModels}
\end{table}
~\\[-7mm]

The design variables are the residence time $u_1 \in [1,100]$, the initial concentration $u_2 \in [0.01,1]$ and the reactor type $u_3 \in \lbrace \mathrm{PFR}, \mathrm{CSTR} \rbrace$. The model parameter is the reaction rate $\theta_i \in [1\textsc{e-}6,\, 0.1]$. 

Note that the expression for the conversion rate in the ideal PFR reactor is the same for micro- and macrofluids. This is also true for the CSTR reactor 1\textsuperscript{st}-order reaction expression. All expression are differentiable with respect to the model parameter, with the exception of the $f_1$ and $f_2$ at $R = 1$.

Experimental data can be generated from any of the models; We propose the following model parameter value for the \new{data-generating} model:\\[-6mm]
\begin{table}[!h]
    \centering
    \begin{tabular}{c c c c c}
         $\theta_1$ & $\theta_2$ & $\theta_3$ & $\theta_4$ & $\theta_5$ \Tstrut\Bstrut \\
         \hline
         6\textsc{e-}3 & 6\textsc{e-}3 & 0.015 & 0.025 & 0.025 \Tstrut\Bstrut
    \end{tabular}
\end{table}
~\\[-7mm]

For the experimental evaluation in this paper, we generate data from model $f_3$ and $f_5$ with experimental noise variance $\sigma^2 = 2.5\textsc{e-}3$. Generating data from model $f_5$ produces a significantly more difficult problem of model discrimination, since models $f_4$ and $f_5$ yield predictions difficult to differentiate.

\section{Case Study: Ammonia Synthesis}
\label{sec:casestudyoneappendix}
This case study looks at four different models for synthesis of ammonia $\mathrm{NH}_{3}$ from hydrogen $\mathrm{H}_{2}$ and nitrogen $\mathrm{N}_{2}$~\citep{BuzziFerraris1990}. Given pressure $P \in [300\,\mathrm{atm},\,350\,\mathrm{atm}]$, temperature $T \in [703\,\mathrm{K},\,753\,\mathrm{K}]$ and inlet ammonia mole fraction $\chi_{\mathrm{NH}_3} \in [0.1,\,0.2]$. 
The models are given by:
\begin{align*}
	\mbox{Model 1}:\quad
    f_1 &= \frac{\phi_{\mathrm{N}_2} - \phi_{\mathrm{NH}_3} / ( \phi_{\mathrm{H}_2}^3 K_\mathrm{eq}^2 ) }{C_1 \phi_{\mathrm{NH}_3} / \phi_{\mathrm{H}_2}^{3/2} } \,, 
    \\[2mm]
	\mbox{Model 2}:\quad
    f_2 &= \frac{\phi_{\mathrm{N}_2} \phi_{\mathrm{H}_2} - \phi_{\mathrm{NH}_3} / (\phi_{\mathrm{H}_2} K_\mathrm{eq})^2}{C_1 \phi_{\mathrm{NH}_3}} \,, 
    \\[2mm]
	\mbox{Model 3}:\quad
    f_3 &= \frac{\phi_{\mathrm{N}_2}^{1/2} \phi_{\mathrm{H}_2}^{3/2} - \phi_{\mathrm{NH}_3} / K_\mathrm{eq} }{C_1 \phi_{\mathrm{NH}_3} + C_2 (\phi_{\mathrm{N}_2} / \phi_{\mathrm{H}_2})^{1/2} } \,, 
    \\[2mm]
	\mbox{Model 4}:\quad
    f_4 &= \frac{\phi_{\mathrm{N}_2}^{1/2} \phi_{\mathrm{H}_2}^{3/2} - \phi_{\mathrm{NH}_3} / K_\mathrm{eq}}{C_1 \phi_{\mathrm{NH}_3} + C_2 \phi_{\mathrm{N}_2} + C_3 \phi_{\mathrm{NH}_3} / \phi_{\mathrm{N}_2} } \,,
\end{align*}
where the fugacities are given by $\phi_{s} = P \chi_{s} \gamma_{s}$ for $s \in \lbrace \mathrm{H}_2,\, \mathrm{N}_2,\, \mathrm{NH}_3 \rbrace$. We assume inert-free, stoichiometric reaction, which gives the mole fractions $\chi_{\mathrm{N}_2} = \tfrac{1}{4} \left( 1 - \chi_{\mathrm{NH}_3} \right)$ and $\chi_{\mathrm{H}_2} = 3 \chi_{\mathrm{N}_2}$. The activity coefficients $\gamma_s$ for the reaction are given by~\citep{DysonSimon1968}:
\begin{align}
    \begin{split}
        \gamma_{\mathrm{H}_2} 
        &= \exp\bigg[ P \exp\left( 0.541 - 3.8402 \cdot T^{0.125} \right)
    	\\&\quad\quad\quad\quad - P^2 \exp \left( -15.98 - 0.1263 \cdot T^{0.5} \right) \
    	\\&\quad\quad\quad\quad+ \frac{300 \left(\exp(-P/300) - 1 \right)}{\exp\left(5.941 + 0.011901 \cdot T \right)} \bigg] \,,
    \end{split}
	\\ %
    \begin{split}
        \gamma_{\mathrm{N}_2} 
	    &= 0.93431737 + 3.101804\textsc{e-}4 \cdot T + 2.958960\textsc{e-}4 \cdot P
	    \\&\quad- 2.707279\textsc{e-}7 \cdot T^2 + 4.775207\textsc{e-}7 \cdot P^2 \,,
    \end{split}
	\\%
    \begin{split}
        \gamma_{\mathrm{NH}_3} 
	    &= 0.14389960 + 2.028538\textsc{e-}3 \cdot T - 4.487672\textsc{e-}4 \cdot P
	    \\&\quad- 1.142945\textsc{e-}6 \cdot T^2 + 2.761216\textsc{e-}7 \cdot P^2 \,.
    \end{split}
\end{align}
The thermodynamic equilibrium constant $K_\mathrm{eq}$ is given by~\citep{GillespieBeattie1930}:
\begin{align}
    \begin{split}
        \log_{10} K_\mathrm{eq} &= 2.6899 - 2.691122 \log_{10}T - 5.519265\textsc{e-}5 \cdot T 
        \\&\quad+ 1.848863\textsc{e-}7 \cdot T^2 + 2001.6 / T \,.
    \end{split}
\end{align}
The model parameters appear in the coefficients $C_j = \exp \left( \theta_{j1} - \theta_{j2} \tfrac{T - 700}{T} \right)$. The bounds on the model parameters are $\theta_{j1} \in [0.1,\,10]$ and $\theta_{j2} \in [0.1,\,100]$.

We follow~\citep{BuzziFerraris1990} by generating experimental data from model 1 with $\vec \theta = [3.68,\, 11.8]$ and experimental noise variance $\vec \Sigma = \sigma^2 = 90$.

\section{Case Study: Chemical Kinetic Models}
\label{sec:casestudytwoappendix}
This case study looks at four chemical kinetic models~\citep{BuzziFerraris1984}. There are two observable outputs $y_1$, $y_2$ and two design variables $u_1,u_2 \in [5,55]$. Each chemical kinetic model $i$ has four model parameters $\theta_{i,j} \in [0,1]$. The model functions are given by:
\begin{alignat*}{2}
	\mbox{Model 1}:\quad f_{1,(1)} &= \theta_{1,1} x_1 x_2/g_1 \,,\quad
    && f_{1,(2)} = \theta_{1,2} x_1 x_2/g_1 \,, \\[2mm]
	\mbox{Model 2}:\quad f_{2,(1)} &= \theta_{2,1} x_1 x_2/g_2^2 \,,\,
    && f_{2,(2)} = \theta_{2,2} x_1 x_2/h_{2,1}^2 \,, \\[2mm]
	\mbox{Model 3}:\quad f_{3,(1)} &= \theta_{3,1} x_1 x_2/h_{3,1}^2 \,, 
    && f_{3,(2)} = \theta_{3,2} x_1 x_2/h_{3,2}^2 \,, \\[2mm]
	\mbox{Model 4}:\quad f_{4,(1)} &= \theta_{4,1} x_1 x_2/g_4 \,, 
    && f_{4,(2)} = \theta_{4,2} x_1 x_2/h_{4,1} \,,
\end{alignat*}
where $g_i = 1 + \theta_{i,3} x_1 + \theta_{i,4} x_2$ and $h_{i,j} = 1 + \theta_{i,2+j} x_j$. We follow~\citep{BuzziFerraris1984} by generating experimental data from model 1 with $\theta_{1,1}=\theta_{1,3}=0.1$ and $\theta_{1,2}=\theta_{1,4}=0.01$ and experimental noise covariance $\vec \Sigma = \diag(0.35,\,2.3\textsc{e-}3)$. We start each test with $5$ randomly sampled experimental observations, and set a maximum budget of $40$ additional experiments.


\section*{References}
\bibliographystyle{elsarticle-harv}\biboptions{authoryear}
\bibliography{ref.bib}

\end{document}